\documentclass[12pt]{article}
\usepackage{amsmath,amssymb}
\usepackage{graphicx}
\newcommand{\eqdef}{\stackrel{\text{def}}{=}}
\newcommand{\n}{\nonumber\\}
\newcommand{\bm}{\boldsymbol}
\newcommand{\cF}{c_{\text{\tiny$\mathcal{F}$}}}
\newcommand{\ignore}[1]{}
\numberwithin{equation}{section}
\newcommand{\Romannumeral}[1]{\uppercase\expandafter{\romannumeral#1}}
\newcommand{\I}{\text{\Romannumeral{1}}}
\newcommand{\II}{\text{\Romannumeral{2}}}
\newcommand{\III}{\text{\Romannumeral{3}}}
\newtheorem{thm}{\bf Theorem}
\newtheorem{prop}{\bf Proposition}

\allowdisplaybreaks[4]

\begin{document}

\baselineskip=20pt

\newfont{\elevenmib}{cmmib10 scaled\magstep1}
\newcommand{\preprint}{
    \begin{flushright}\normalsize \sf
     DPSU-15-1\\
   \end{flushright}}
\newcommand{\Title}[1]{{\baselineskip=26pt
   \begin{center} \Large \bf #1 \\ \ \\ \end{center}}}
\newcommand{\Author}{\begin{center}
   \large \bf Satoru Odake \end{center}}
\newcommand{\Address}{\begin{center}
     Faculty of Science, Shinshu University,\\
     Matsumoto 390-8621, Japan
   \end{center}}
\newcommand{\Accepted}[1]{\begin{center}
   {\large \sf #1}\\ \vspace{1mm}{\small \sf Accepted for Publication}
   \end{center}}

\preprint
\thispagestyle{empty}

\Title{Recurrence Relations of\\ the Multi-Indexed Orthogonal Polynomials
: $\III$}

\Author

\Address
\vspace{1cm}

\begin{abstract}
In a previous paper, we presented conjectures of the recurrence relations
with constant coefficients for the multi-indexed orthogonal polynomials of
Laguerre, Jacobi, Wilson and Askey-Wilson types.
In this paper we present a proof for the Laguerre and Jacobi cases.
Their bispectral properties are also discussed, which give a method to
obtain the coefficients of the recurrence relations explicitly.
This paper extends to the Laguerre and Jacobi cases the bispectral
techniques recently introduced by G\'{o}mez-Ullate et al. to derive
explicit expressions for the coefficients of the recurrence relations
satisfied by exceptional polynomials of Hermite type.
\end{abstract}

\section{Introduction}
\label{intro}

The exceptional orthogonal polynomials have seen remarkable developments in
recent years in connection with exactly solvable quantum mechanical systems
in one dimension \cite{gkm08}--\cite{llms14} (and the references therein).
The exceptional orthogonal polynomials
$\{\mathcal{P}_n(\eta)|n\in\mathbb{Z}_{\geq 0}\}$ satisfy second order
differential or difference equations and form a complete set, but
there are missing degrees, by which the constraints of Bochner's theorem
and its generalizations \cite{bochner,szego} are avoided.
We distinguish the following two cases;
the set of missing degrees $\mathcal{I}=\mathbb{Z}_{\geq 0}\backslash
\{\text{deg}\,\mathcal{P}_n|n\in\mathbb{Z}_{\geq 0}\}$ is
case (1): $\mathcal{I}=\{0,1,\ldots,\ell-1\}$, or 
case (2) $\mathcal{I}\neq\{0,1,\ldots,\ell-1\}$, where $\ell$ is a positive
integer. The situation of case (1) is called stable in \cite{gkm11}.
By applying the multi-step Darboux transformation \cite{darb} to the
quantum mechanical systems described by the classical orthogonal polynomials,
various exceptional orthogonal polynomials with multi-indices can be obtained.
The choice of the seed solutions of the Darboux transformation leads to
case (1) or case (2).
When the eigenstate or pseudo virtual state wavefunctions are used as seed
solutions, we obtain case (2) \cite{os29,os30}.
When the virtual state wavefunctions are used as seed solutions, we obtain
case (1) and call them multi-indexed orthogonal polynomials
\cite{os25,os27,os26}.

The ordinary orthogonal polynomials
$\{P_n(\eta)|n\in\mathbb{Z}_{\geq 0},\text{deg}\,P_n=n\}$
satisfy the three term recurrence relations, and conversely the polynomials
satisfying the three term recurrence relations are orthogonal polynomials
(Favard's theorem \cite{szego}).
Since the exceptional orthogonal polynomials are not ordinary orthogonal
polynomials, they do not satisfy the three term recurrence relations.
Recurrence relations for exceptional polynomials were discussed by several
authors \cite{stz10,rrmiop,ggm13,d14,mt14,rrmiop2}.
In our first paper \cite{rrmiop}, we showed that $M$-indexed orthogonal
polynomials $P_{\mathcal{D},n}(\eta)$ ($\mathcal{D}=\{d_1,\ldots,d_M\}$)
of Laguerre, Jacobi, Wilson and Askey-Wilson types satisfy $3+2M$ term
recurrence relations with variable dependent coefficients.
In our second paper \cite{rrmiop2}, we discussed recurrence relations with
constant coefficients for the multi-indexed orthogonal polynomials of
Laguerre, Jacobi, Wilson and Askey-Wilson types,
$X(\eta)P_{\mathcal{D},n}(\eta)
=\sum\limits_{k=-L}^Lr_{n,k}^{X,\mathcal{D}}P_{\mathcal{D},n+k}(\eta)$,
and gave conjectures on the condition for the polynomial $X(\eta)$.
Recently G\'{o}mez-Ullate, Kasman, Kuijlaars and Milson studied the
exceptional Hermite polynomials with multi-indices and showed the recurrence
relations with constant coefficients \cite{gkkm15}.
Their method can be applied to the Laguerre and Jacobi cases and we can prove
the recurrence relations with constant coefficients for the multi-index
Laguerre and Jacobi polynomials conjectured in \cite{rrmiop2}.
This is the first motivation of the present paper.

The second motivation of the present paper is a study of the bispectral
property \cite{szego,dg86}:
\begin{equation}
  \widetilde{\mathcal{H}}_{\mathcal{D}}P_{\mathcal{D},n}(\eta)
  =\mathcal{E}_nP_{\mathcal{D},n}(\eta),\quad
  \Delta_{X,\mathcal{D}}P_{\mathcal{D},n}(\eta)
  =X(\eta)P_{\mathcal{D},n}(\eta),  
  \label{bispec}
\end{equation}
where $\widetilde{\mathcal{H}}_{\mathcal{D}}$ is the second order
differential operator of $\eta$ and $\Delta_{X,\mathcal{D}}$ is a certain
shift operator of $n$.
In \cite{gkkm15} they also studied bispectral properties of the exceptional
Hermite polynomials with multi-indices.
Their key point is the anti-isomorphism $\flat$, which originates from
`bispectral Darboux transformation' \cite{kr97}.
We explain it briefly.
The operators $\partial_{\eta}$ and $\eta$ act on the Hermite polynomial
$H_n(\eta)$ as $\partial_{\eta}H_n(\eta)=2nH_{n-1}(\eta)$ and
$\eta H_n(\eta)=\frac12H_{n+1}(\eta)+nH_{n-1}(\eta)$.
By introducing the operators $\Gamma=2ne^{-\partial_n}$ and
$\Delta=\frac12e^{\partial_n}+ne^{-\partial_n}$, we have
$\partial_{\eta}H_n(\eta)=\Gamma H_n(\eta)$ and
$\eta H_n(\eta)=\Delta H_n(\eta)$.
Since commutators among these operators are
$[\partial_{\eta},\eta]=1$ and $[\Delta,\Gamma]=1$ (and other commutators
vanish), we have an algebra anti-isomorphism
$\flat:\mathbb{C}[\partial_{\eta},\eta]\to\mathbb{C}[\Delta,\Gamma]$,
$\flat(\eta^i\partial_{\eta}^j)=\Gamma^j\Delta^i$ ($i,j=0,1,\ldots$).
The exceptional Hermite polynomial $P_{\mathcal{D},n}(\eta)$ and the original
Hermite polynomial $H_n(\eta)$ are related by the multi-step forward and
backward shift operators, $\hat{\mathcal{F}}^{(\mathcal{D})}$ and
$\hat{\mathcal{B}}^{(\mathcal{D})}$
(These are our notation, see Appendix\,\ref{app:DT}.
$\hat{\mathcal{F}}^{(\mathcal{D})}$, $\hat{\mathcal{B}}^{(\mathcal{D})}$
and $\eta$ correspond to $A$, $B$ and $x$ in \cite{gkkm15}, respectively).
They are differential operators of $\eta$
($\hat{\mathcal{F}}^{(\mathcal{D})}\in\mathbb{C}[\partial_{\eta},\eta]$,
$\hat{\mathcal{B}}^{(\mathcal{D})}\not\in\mathbb{C}[\partial_{\eta},\eta]$)
and commute with $\Delta$ and $\Gamma$.
For an appropriate polynomial $X(\eta)$ that gives recurrence relations with
constant coefficients, the operator
$\Theta_{X,\mathcal{D}}=\hat{\mathcal{B}}^{(\mathcal{D})}\circ
X(\eta)\circ\hat{\mathcal{F}}^{(\mathcal{D})}$ belongs to
$\mathbb{C}[\partial_{\eta},\eta]$. Then the operator
$\Delta_{X,\mathcal{D}}=\flat(\Theta_{X,\mathcal{D}})
\circ\pi_{\mathcal{D}}^{-1}(n)$, where $\pi_{\mathcal{D}}(n)$ is a certain
function of $n$ and $f^{-1}$ means $f^{-1}(x)=f(x)^{-1}$, gives
$X(\eta)P_{\mathcal{D},n}(\eta)=\Delta_{X,\mathcal{D}}P_{\mathcal{D},n}(\eta)$
($X$ and $\Delta_{X,\mathcal{D}}$ correspond to $f$, $\tilde{\Delta}_f$
in \cite{gkkm15}, respectively).
To derive this result, the commutativity
$[\Delta_{X,\mathcal{D}},\hat{\mathcal{B}}^{(\mathcal{D})}]=0$ is important.
By using this result, we can obtain the coefficients $r_{n,k}^{X,\mathcal{D}}$
explicitly.

This argument can be applied to the Laguerre and Jacobi cases but a slight
modification is needed. The reason is that the Hermite polynomial $H_n(\eta)$
has no parameter but the Laguerre $L^{(\alpha)}_n(\eta)$ and Jacobi
$P^{(\alpha,\beta)}_n(\eta)$ polynomials have parameters ($\alpha$ and
$\beta$). We explain this taking the Laguerre case as an example.
The three term recurrence relations of the Laguerre polynomial
$L^{(\alpha)}_n(\eta)$ give
$\eta L^{(\alpha)}_n(\eta)=\Delta L^{(\alpha)}_n(\eta)$,
$\Delta=-(n+1)e^{\partial_n}+2n+\alpha+1-(n+\alpha)e^{-\partial_n}$.
For differentiation, a well known formula is the forward shift relation
$\partial_{\eta}L^{(\alpha)}(\eta)=-L^{(\alpha+1)}_{n-1}(\eta)$ and it
may lead us to define $\Gamma'=-e^{-\partial_n}e^{\partial_{\alpha}}$.
Their commutators are
\begin{equation}
  [\Delta,\Gamma']=I',\quad[I',\Delta]=[I',\Gamma']=0,\quad
  I'=(1-e^{-\partial_n})e^{\partial_{\alpha}},\quad
  I'L^{(\alpha)}_n(\eta)=L^{(\alpha)}_n(\eta),
\end{equation}
and $\Delta$, $\Gamma'$ and $I'$ commute with $\partial_{\eta}$ and $\eta$.
However $\Gamma'$ and $I'$ do not commute with $\alpha$.
Since the operator $\hat{\mathcal{B}}^{(\mathcal{D})}$ contains the parameter
$\alpha$ as a coefficient of $\partial_{\eta}^k$, the commutativity
$[\Delta_{X,\mathcal{D}},\hat{\mathcal{B}}^{(\mathcal{D})}]=0$ is lost.
The operator $\Gamma$ should contain $n$-shifts only.
The expression of $\Gamma$ becomes more complicated than the Hermite case.
The important map $\flat$ can be defined but it is no longer anti-isomorphism.
The details are given in the main text.

This paper is organized as follows.
In section \ref{sec:RR} we prove the conjecture of the recurrence relations
with constant coefficients for the multi-indexed Laguerre and Jacobi
polynomials.
After recapitulating some fundamental formulas of the multi-indexed Laguerre
and Jacobi polynomials in \S\,\ref{sec:miop} and the conjecture in
\S\,\ref{sec:rr}, a proof is given in \S\,\ref{sec:proof}.
In section \ref{sec:BP} we discuss the bispectral property of the multi-indexed
Laguerre and Jacobi polynomials. After preparing some algebra and shift
operators in
\S\,\ref{sec:prepare}, we define the map $\flat$ for any ordinary
orthogonal polynomials in continuous variable in \S\,\ref{sec:mapb}.
By using this map, the bispectral property,
Theorem\,\ref{thmbispec}, is established in \S\,\ref{sec:bispec}. Examples
for Theorem\,\ref{thmbispec} are presented in \S\,\ref{sec:ex}.
The final section is for a summary and comments.
In Appendix\,\ref{app:DT} we review the algebraic aspects of the Darboux
transformation, which are used to derive various properties of the exceptional
orthogonal polynomials with multi-indices.
In Appendix\,\ref{app:JL} the algebraic properties of the multi-indexed
Laguerre and Jacobi orthogonal polynomials are reviewed.
The formulas \eqref{Ahdsphisn=W/W} and \eqref{Bh(s)Psn} are new.
These two Appendices fix the notation in this paper.

\section{Recurrence Relations with Constant Coefficients}
\label{sec:RR}

In this section we prove the conjecture of the recurrence relations with
constant coefficients for multi-indexed Laguerre and Jacobi orthogonal
polynomials given in \cite{rrmiop2}.

\subsection{Multi-indexed orthogonal polynomials}
\label{sec:miop}

The Darboux transformation and the multi-indexed orthogonal polynomials
of Laguerre and Jacobi types are reviewed in Appendix\,\ref{app:DT} and
\ref{app:JL}, and we follow the notation there.
For a set of labels $\mathcal{D}=\{d_1,\ldots,d_M\}$, we write
$\mathcal{H}_{d_1\ldots d_M}$, $\phi_{d_1\ldots d_M\,n}(x)$,
$P_{d_1\ldots d_M,n}(\eta)$, $\Xi_{d_1\ldots d_M}(\eta)$,
$\hat{\mathcal{A}}_{d_1\ldots d_M}$, $\hat{\mathcal{A}}^{(d_1\ldots d_M)}$,
$\hat{\mathcal{F}}_{d_1\ldots d_M}$, $\hat{\mathcal{F}}^{(d_1\ldots d_M)}$,
$\ell_{d_1\ldots d_M}$,
etc.\! as
$\mathcal{H}_{\mathcal{D}}$, $\phi_{\mathcal{D}\,n}(x)$,
$P_{\mathcal{D},n}(\eta)$, $\Xi_{\mathcal{D}}(\eta)$,
$\hat{\mathcal{A}}_{\mathcal{D}}$, $\hat{\mathcal{A}}^{(\mathcal{D})}$,
$\hat{\mathcal{F}}_{\mathcal{D}}$, $\hat{\mathcal{F}}^{(\mathcal{D})}$,
$\ell_{\mathcal{D}}$,
etc., respectively.
We assume that the parameters ($g$ and $h$) are generic such that
$c_{\mathcal{D}}^{\Xi}\neq 0$ \eqref{cXis},
$c_{\mathcal{D},n}^{P}\neq 0$ \eqref{cPsn} and
$\mathcal{E}_n-\tilde{\mathcal{E}}_{d_j}\neq 0$.

The multi-indexed orthogonal polynomials of the Laguerre and Jacobi types
$P_{\mathcal{D},n}(\eta)$ and the original Laguerre and Jacobi polynomials
$P_n(\eta)$ are related as follows:
\begin{align}
  \hat{\mathcal{F}}^{(\mathcal{D})}P_n(\eta)
  &=\rho^{(\mathcal{D})}_{\hat{\mathcal{F}}}(\eta)
  \text{W}[\mu_{d_1},\ldots,\mu_{d_M},P_n](\eta)
  =P_{\mathcal{D},n}(\eta),
  \label{FhDPn=}\\
  \hat{\mathcal{B}}^{(\mathcal{D})}P_{\mathcal{D},n}(\eta)
  &=\rho^{(\mathcal{D})}_{\hat{\mathcal{B}}}(\eta)
  \text{W}[m_1,\ldots,m_M,P_n](\eta)
  =\pi_{\mathcal{D}}(n)P_n(\eta),
  \label{BhDPDn=}
\end{align}
where $\hat{\mathcal{F}}^{(\mathcal{D})}$,
$\hat{\mathcal{B}}^{(\mathcal{D})}$, $\mu_{\text{v}}(\eta)$,
$\rho^{(\mathcal{D})}_{\hat{\mathcal{F}}}(\eta)$,
$\rho^{(\mathcal{D})}_{\hat{\mathcal{B}}}(\eta)$ and
$m_j(\eta)=m^{(\mathcal{D})}_j(\eta)$ are defined by 
\eqref{Fh(s)}, \eqref{Bh(s)}, \eqref{muv}, \eqref{rhoFh},
\eqref{rhoBh} and \eqref{mj}, respectively (see also \eqref{FhsPn=}),
and the constant $\pi_{\mathcal{D}}(n)$ is defined by
\begin{equation}
  \pi_{\mathcal{D}}(n)\eqdef\prod_{j=1}^M
  (\mathcal{E}_n-\tilde{\mathcal{E}}_{d_j}).
  \label{piD}
\end{equation}
This polynomial $P_{\mathcal{D},n}(\eta)$ satisfies the second order
differential equation (see \eqref{c1c2}--\eqref{HtsPsn}),
\begin{gather}
  \widetilde{\mathcal{H}}_{\mathcal{D}}P_{\mathcal{D},n}(\eta)
  =\mathcal{E}_nP_{\mathcal{D},n}(\eta),
  \label{HtDPDn}\\
  -\tfrac14\widetilde{\mathcal{H}}_{\mathcal{D}}
  =c_2(\eta)\frac{d^2}{d\eta^2}
  +\Bigl(c_{11}(\eta)-2c_2(\eta)
  \frac{\partial_{\eta}\Xi_{\mathcal{D}}(\eta)}
  {\Xi_{\mathcal{D}}(\eta)}\Bigr)\frac{d}{d\eta}
  +c_2(\eta)
  \frac{\partial^2_{\eta}\Xi_{\mathcal{D}}(\eta)}
  {\Xi_{\mathcal{D}}(\eta)}
  -c_{10}(\eta)
  \frac{\partial_{\eta}\Xi_{\mathcal{D}}(\eta)}
  {\Xi_{\mathcal{D}}(\eta)}.
  \label{HtD}
\end{gather}
Here $c_{11}(\eta)=c_1(\eta,\bm{\lambda}^{[M_{\I},M_{\II}]})$,
$c_{10}(\eta)=c_1(\eta,\bm{\lambda}^{[M_{\I},M_{\II}]}-\bm{\delta})$ and
$c_2(\eta)$ are
\begin{align}
  c_{11}(\eta)&=\left\{
  \begin{array}{ll}
  g+M_{\I}-M_{\II}+\frac12-\eta&:\text{L}\\
  h-g-2M_{\I}+2M_{\II}-(g+h+1)\eta&:\text{J}
  \end{array}\right.,\\
  c_{10}(\eta)&=c_{11}(\eta)+\left\{
  \begin{array}{ll}
  -1&:\text{L}\\
  2\eta&:\text{J}
  \end{array}\right.,\quad
  c_2(\eta)=\left\{
  \begin{array}{ll}
  \eta&:\text{L}\\
  1-\eta^2&:\text{J}
  \end{array}\right.,
\end{align}
where
$M_{\text{t}}=\#\{d_j\,|\,d_j\text{\,:\ type $\text{t}$},\,j=1,\ldots,M\}$
($\text{t}=\I,\II$).
The degrees of $P_{\mathcal{D},n}(\eta)$ and $\Xi_{\mathcal{D}}(\eta)$
are $\ell_{\mathcal{D}}+n$ and $\ell_{\mathcal{D}}$ respectively, and
$\ell_{\mathcal{D}}$ is given in \eqref{ells}.
We set $P_n(\eta)=P_{\mathcal{D},n}(\eta)=0$ for $n<0$.

\subsection{Recurrence relations with constant coefficients}
\label{sec:rr}

In our previous paper \cite{rrmiop2}, we discussed the recurrence relations
of the multi-indexed Laguerre or Jacobi polynomials
with constant coefficients,
\begin{equation}
 X(\eta)P_{\mathcal{D},n}(\eta)
 =\sum_{k=-L}^Lr_{n,k}^{X,\mathcal{D}}P_{\mathcal{D},n+k}(\eta)
 \ \ (\forall n\in\mathbb{Z}_{\geq 0}),
 \label{XP}
\end{equation}
where $r_{n,k}^{X,\mathcal{D}}\,$'s are constants and $X(\eta)$ is some
polynomial of degree $L$ in $\eta$.
To find such $X(\eta)$ is our purpose.
This problem is rephrased as follows (Remark 3 in \S\,$\II$ of
\cite{rrmiop2}): Find a polynomial $X(\eta)$ such that the operator
$\Theta_{X,\mathcal{D}}\eqdef\hat{\mathcal{B}}^{(\mathcal{D})}\circ
X(\eta)\circ\hat{\mathcal{F}}^{(\mathcal{D})}$
maps polynomials in $\eta$ to polynomials in $\eta$.
The coefficients $r_{n,k}^{X,\mathcal{D}}$ are expressed as
(Proposition 1 in \cite{rrmiop2})
\begin{equation}
  r_{n,k}^{X,\mathcal{D}}=\frac{r_{n,k}^{(0)\,X,\mathcal{D}}}
  {\prod_{j=1}^M(\mathcal{E}_{n+k}-\tilde{\mathcal{E}}_{d_j})},
  \label{rnk=rnk0/E}
\end{equation}
where the constants $r_{n,k}^{(0)\,X,\mathcal{D}}$ are obtained from
the relations among the classical orthogonal polynomials
\begin{equation}
  \Theta_{X,\mathcal{D}}P_n(\eta)
  =\sum_{k=-n}^Lr_{n,k}^{(0)\,X,\mathcal{D}}P_{n+k}(\eta)
  \ \Bigl(=\sum_{k=-L}^Lr_{n,k}^{(0)\,X,\mathcal{D}}P_{n+k}(\eta)\Bigr).
  \label{rnk0}
\end{equation}

If the two polynomials in $\eta$,
$\Xi_{\mathcal{D}}(\eta)=\Xi_{d_1\ldots d_M}(\eta)$ and
$\Xi_{d_1\ldots d_{M-1}}(\eta)$, do not have common roots,
the necessary condition for $X(\eta)$ is the following
(Proposition 2 and its Remark in \cite{rrmiop2}):
$\frac{dX(\eta)}{d\eta}$ is divisible by $\Xi_{\mathcal{D}}(\eta)$, namely
\begin{equation}
  \frac{dX(\eta)}{d\eta}
  =\Xi_{\mathcal{D}}(\eta)Y(\eta),\quad
  \text{\rm $Y(\eta)$ : a polynomial in $\eta$}.
  \label{Xcond}
\end{equation}
Since the overall normalization and the constant term of $X(\eta)$ are
irrelevant, we take the candidate $X(\eta)$ as
\begin{equation}
  X(\eta)=\int_0^{\eta}\Xi_{\mathcal{D}}(y)Y(y)dy,\quad
  \text{deg}\,X(\eta)=L=\ell_{\mathcal{D}}+\text{deg}\,Y(\eta)+1,
  \label{X=int}
\end{equation}
and we assume $Y(\eta)\in\mathbb{C}[\eta,g,h]$.
The Conjecture given in \cite{rrmiop2} is that the polynomial $X(\eta)$
satisfying \eqref{Xcond} gives \eqref{XP}.
Since we will prove this conjecture in the next subsection, we state it as
a theorem:
\begin{thm}
For any polynomial $Y(\eta)$, we define $X(\eta)$ as \eqref{X=int}. Then the
multi-indexed Laguerre and Jacobi polynomials $P_{\mathcal{D},n}(\eta)$
satisfy $1+2L$ term recurrence relations with constant coefficients \eqref{XP}.
{\rm (}See Remark in \S\,\ref{sec:proof}.{\rm )}
\label{conj_oQM}
\end{thm}
{\bf Remark 1}$\,$
If two polynomials in $\eta$,
$\Xi_{\mathcal{D}}(\eta)=\Xi_{d_1\ldots d_M}(\eta)$ and
$\Xi_{d_1\ldots d_{M-1}}(\eta)$, do not have common roots,
this theorem exhausts all possible $X(\eta)$ giving recurrence relations
with constant coefficients \cite{rrmiop2}.\\
{\bf Remark 2}$\,$
If $\frac{dX(\eta)}{d\eta}$ is divisible by $\Xi_{\mathcal{D}}(\eta)$,
we have $\Theta_{X,\mathcal{D}}\in\mathbb{C}[\partial_{\eta},\eta]$.

Some examples for \eqref{XP} are found in \cite{stz10,d14,mt14,rrmiop2}.

\subsection{Proof}
\label{sec:proof}

Following the argument in \cite{gkkm15}, we prove Theorem\,\ref{conj_oQM}.

Let us define the set of finite linear combinations of
$P_{\mathcal{D},n}(\eta)$, $\mathcal{U}_{\mathcal{D}}\subset\mathbb{C}[\eta]$,
and the stabilizer ring $\mathcal{S}_{\mathcal{D}}\subset\mathbb{C}[\eta]$ by
\begin{align}
  \mathcal{U}_{\mathcal{D}}
  &\eqdef\text{Span}\{P_{\mathcal{D},n}(\eta)\bigm|n\in\mathbb{Z}_{n\geq 0}\},
  \label{UD}\\
  \mathcal{S}_{\mathcal{D}}
  &\eqdef\bigl\{X(\eta)\in\mathbb{C}[\eta]\bigm|
  X(\eta)P_{\mathcal{D},n}(\eta)\in\mathcal{U}_{\mathcal{D}}
  \ \ (\forall n\in\mathbb{Z}_{\geq0})\bigr\}.
  \label{SD}
\end{align}
Since the degree of $P_{\mathcal{D},n}(\eta)$ is $\ell_{\mathcal{D}}+n$,
it is trivial that $p(\eta)\in\mathcal{U}_{\mathcal{D}}\Rightarrow
\deg\,p\geq\ell_{\mathcal{D}}$, except for $p(\eta)=0$.

For $p(\eta)\in\mathcal{U}_{\mathcal{D}}$, let us expand it as
$p(\eta)=\!\!\!\sum\limits_{n=0}^{\deg\,p-\ell_{\mathcal{D}}}\!\!\!
a_nP_{\mathcal{D},n}(\eta)$ ($a_n$: constant)
and consider the action of $\widetilde{\mathcal{H}}_{\mathcal{D}}$ on it.
{}From \eqref{HtDPDn}, we have
$\widetilde{\mathcal{H}}_{\mathcal{D}}\,p(\eta)
=\!\!\!\sum\limits_{n=0}^{\deg\,p-\ell_{\mathcal{D}}}\!\!\!
a_n\mathcal{E}_nP_{\mathcal{D},n}(\eta)\in\mathbb{C}[\eta]$.
On the other hand, from \eqref{HtD}, we have
\begin{align}
  \widetilde{\mathcal{H}}_{\mathcal{D}}\,p(\eta)
  &=-4\bigl(c_2(\eta)\partial_{\eta}^2p(\eta)
  +c_{11}(\eta)\partial_{\eta}p(\eta)\bigr)\n
  &\quad+\frac{4}{\Xi_{\mathcal{D}}(\eta)}\Bigl(
  \partial_{\eta}\Xi_{\mathcal{D}}(\eta)\bigl(
  2c_2(\eta)\partial_{\eta}p(\eta)+c_{10}(\eta)p(\eta)\bigr)
  -\partial_{\eta}^2\Xi_{\mathcal{D}}(\eta)c_2(\eta)p(\eta)\Bigr).
\end{align}
Since the first line of r.h.s is a polynomial in $\eta$, we obtain
the condition:
\begin{equation}
  \partial_{\eta}\Xi_{\mathcal{D}}(\eta)\bigl(
  2c_2(\eta)\partial_{\eta}p(\eta)+c_{10}(\eta)p(\eta)\bigr)
  -\partial_{\eta}^2\Xi_{\mathcal{D}}(\eta)c_2(\eta)p(\eta)
  \text{ is divisible by }\Xi_{\mathcal{D}}(\eta).
  \label{pcond}
\end{equation}
Next let us consider the converse. Take any polynomial $p(\eta)$ satisfying
the condition \eqref{pcond} and expand it as
\begin{equation}
  p(\eta)=\!\!\!\sum_{n=0}^{\deg\,p-\ell_{\mathcal{D}}}\!\!\!
  a_nP_{\mathcal{D},n}(\eta)+r(\eta),\quad
  \deg\,r<\ell_{\mathcal{D}},\quad
  r(\eta)=\sum_{k=0}^{\deg\,r}r_k\eta^k,
\end{equation}
($p(\eta)=r(\eta)$ for $\deg\,p<\ell_{\mathcal{D}}$).
Since $P_{\mathcal{D},n}(\eta)$ satisfies \eqref{pcond}, the condition
\eqref{pcond} becomes
\begin{equation}
  \partial_{\eta}\Xi_{\mathcal{D}}(\eta)\bigl(
  2c_2(\eta)\partial_{\eta}r(\eta)+c_{10}(\eta)r(\eta)\bigr)
  -\partial_{\eta}^2\Xi_{\mathcal{D}}(\eta)c_2(\eta)r(\eta)
  \text{ is divisible by }\Xi_{\mathcal{D}}(\eta).
  \label{rcond}
\end{equation}
In general the polynomial $\Xi_{\mathcal{D}}(\eta)$ has only simple zeros,
$\Xi_{\mathcal{D}}(\eta)\propto
\prod\limits_{i=1}^{\ell_{\mathcal{D}}}(\eta-\eta_i)$.
The condition \eqref{rcond} means that the polynomial in \eqref{rcond}
vanishes at $\eta=\eta_i$. This gives $\ell_{\mathcal{D}}$ linear relations
on $r_k$'s. Since these linear relations are independent and
the number of $r_k$'s is $\deg\,r+1\leq\ell_{\mathcal{D}}$,
all $r_k$'s vanish. Namely we obtain $r(\eta)=0$ and
$p(\eta)\in\mathcal{U}_{\mathcal{D}}$.
We remark that the polynomials $p(\eta)$ satisfying the condition
\eqref{pcond} form a vector space and its codimension in $\mathbb{C}[\eta]$
is $\ell_{\mathcal{D}}$ for $r(\eta)=0$ case.
We summarize this argument as the following proposition.
\begin{prop}
When $\Xi_{\mathcal{D}}(\eta)$ has only simple zeros,
a polynomial $p(\eta)$ belongs to $\mathcal{U}_{\mathcal{D}}$
if and only if $p(\eta)$ satisfies the condition \eqref{pcond}.
\label{ch_UD}
\end{prop}

For any polynomial $X(\eta)$ and $p(\eta)$, we set $q(\eta)=X(\eta)p(\eta)$.
Then the condition \eqref{pcond} for $q(\eta)$ becomes
\begin{align}
  &\quad\partial_{\eta}\Xi_{\mathcal{D}}(\eta)\bigl(
  2c_2(\eta)\partial_{\eta}q(\eta)+c_{10}(\eta)q(\eta)\bigr)
  -\partial_{\eta}^2\Xi_{\mathcal{D}}(\eta)c_2(\eta)q(\eta)\n
  &=X(\eta)\Bigl(\partial_{\eta}\Xi_{\mathcal{D}}(\eta)\bigl(
  2c_2(\eta)\partial_{\eta}p(\eta)+c_{10}(\eta)p(\eta)\bigr)
  -\partial_{\eta}^2\Xi_{\mathcal{D}}(\eta)c_2(\eta)p(\eta)\Bigr)\n
  &\quad+\partial_{\eta}\Xi_{\mathcal{D}}(\eta)2c_2(\eta)
  \partial_{\eta}X(\eta)p(\eta).
  \label{qcond}
\end{align}
If $X(\eta)$ satisfies \eqref{Xcond} and $p(\eta)$ belongs to
$\mathcal{U}_{\mathcal{D}}$, this is divisible by $\Xi_{\mathcal{D}}(\eta)$.
When $\Xi_{\mathcal{D}}(\eta)$ has only simple zeros,
Proposition\,\ref{ch_UD} implies $q(\eta)\in\mathcal{U}_{\mathcal{D}}$.
Thus we obtain $X(\eta)\in\mathcal{S}_{\mathcal{D}}$, namely,
the relation among the polynomials \eqref{XP}.
The denominator polynomial $\Xi_{\mathcal{D}}(\eta)$ contains a set of
parameters $\bm{\lambda}$ ($\bm{\lambda}=g$ for Laguerre and
$\bm{\lambda}=(g,h)$ for Jacobi) and it could be made to have higher order
zeros by tuning $\bm{\lambda}$.
Such tuning, however, does not cause any trouble to the relation among
the polynomials \eqref{XP}.
This is shown as follows.
The denominator polynomial $\Xi_{\mathcal{D}}(\eta)$ belongs to
$\mathbb{C}[\eta,g,h]$. The polynomial $X(\eta)$ \eqref{X=int} also belongs
to $\mathbb{C}[\eta,g,h]$ because we assume $Y(\eta)\in\mathbb{C}[\eta,g,h]$.
{}From \eqref{FhD} with \eqref{muv} and \eqref{rhoFh},
$\hat{\mathcal{F}}^{(\mathcal{D})}$ belongs to
$\mathbb{C}[\partial_{\eta},\eta,g,h]$.
{}From \eqref{FhD} with \eqref{rhoBh}--\eqref{mj}, the coefficients of
$\partial_{\eta}^k$'s in $\hat{\mathcal{B}}^{(\mathcal{D})}$ are rational
functions of $\eta$ and the factor in the denominator is only
$\Xi_{\mathcal{D}}(\eta)^M$. This factor is factorized as
$\Xi_{\mathcal{D}}(\eta)=c^{\Xi}_{\mathcal{D}}
\prod\limits_{i=1}^{\ell_{\mathcal{D}}}(\eta-\eta_i)$.
Since we already know
$\Theta_{X,\mathcal{D}}=\hat{\mathcal{B}}^{(\mathcal{D})}\circ X(\eta)\circ
\hat{\mathcal{F}}^{(\mathcal{D})}\in\mathbb{C}[\partial_{\eta},\eta]$,
this factor $(\eta-\eta_i)$ is canceled out in $\Theta_{X,\mathcal{D}}$. 
Thus the factor in the denominator of $\Theta_{X,\mathcal{D}}$ is only
$c^{\Xi}_{\mathcal{D}}$. By our assumption, this $c^{\Xi}_{\mathcal{D}}$
does not vanish. 
Therefore \eqref{XP} is valid even when $\Xi_{\mathcal{D}}(\eta)$ has
higher order zeros.
Thus Theorem\,\ref{conj_oQM} is proved.

\noindent
{\bf Remark}$\,$
If $\Xi_{\mathcal{D}}(\eta)$ has only simple zeros, the converse of
Theorem\,\ref{conj_oQM} holds.
To show this, assume that $X(\eta)\in\mathcal{S}_{\mathcal{D}}$,
$p(\eta)\in\mathcal{U}_{\mathcal{D}}$ and \eqref{qcond} is divisible by
$\Xi_{\mathcal{D}}(\eta)$.
Since the expression in the second line of \eqref{qcond} is divisible
by $\Xi_{\mathcal{D}}(\eta)$, the expression in the third line should
be divisible by $\Xi_{\mathcal{D}}(\eta)$.
Since $p(\eta)$ is arbitrary,
$\partial_{\eta}\Xi_{\mathcal{D}}(\eta)c_2(\eta)\partial_{\eta}X(\eta)$
should be divisible by $\Xi_{\mathcal{D}}(\eta)$.
If $\Xi_{\mathcal{D}}(\eta)$ and $\partial_{\eta}\Xi_{\mathcal{D}}(\eta)$
do note have common roots, which happens if $\Xi_{\mathcal{D}}(\eta)$ has
only simple zeros, $\partial_{\eta}X(\eta)$ should be divisible by
$\Xi_{\mathcal{D}}(\eta)$.

We present examples of $\Xi_{\mathcal{D}}(\eta)$ which has higher order
zeros \cite{st12}. We take $\mathcal{D}=\{1^{\I},2^{\II}\}$.
For the Laguerre case, the denominator polynomial is
\begin{equation}
  -2\Xi_{\mathcal{D}}(\eta)
  =\eta^4+2(2g-3)\eta^3+(g-\tfrac52)(6g-1)\eta^2+2(g-\tfrac52)_2(2g+1)\eta
  +(g-\tfrac52)_4,
\end{equation}
which has higher order zeros for $g=-\frac12,\frac32,\frac52,-\frac{13}{2}$.
For these values, $-2\Xi_{\mathcal{D}}(\eta)$ is
$\eta^2(\eta-2)(\eta-6)$, $\eta^2(\eta^2-8)$, $\eta^3(\eta+4)$ and
$(\eta-6)^3(\eta-14)$, respectively.
We can check that $8\Theta_{X,\mathcal{D}}$ belongs to
$\mathbb{Z}[\partial_{\eta},\eta,g]$ and nothing happens at
$g=-\frac12,\frac32,\frac52,-\frac{13}{2}$.
For the Jacobi case, the denominator polynomial is ($a=g+h$, $b=g-h$)
\begin{align}
  64\Xi_{\mathcal{D}}(\eta)
  &=(b-4)(b-3)(b-1)(b+2)\eta^4
  +4(a-1)(b-3)(b-1)b\eta^3\n
  &\quad +2(b-1)\bigl(a(a-2)(3b-4)+(b+4)(b-3)\bigr)\eta^2\n
  &\quad +4(a-1)(b-1)\bigl(a(a-2)+b-3\bigr)\eta\n
  &\quad +a^3(a-4)+2a^2(b-3)-4a(b-5)
  -(b-3)(b-1),
\end{align}
which has higher order zeros for $g=-\frac12,\frac32,\frac52$,
or $h=-\frac32,-\frac12,\frac32$, or 
$(g+h)(g+h-2)(g-h-28)=(g-h-3)(g-h-1)(g-h+4)$
(The cases $g-h=4,3,1,-2$ are excluded by the condition
$c^{\Xi}_{\mathcal{D}}\neq 0$.).
We can check that $16\Theta_{X,\mathcal{D}}$ belongs to
$\mathbb{Z}[\partial_{\eta},\eta,g,h]$ and nothing happens at
these values.

\section{Bispectral Property}
\label{sec:BP}

In this section we discuss the bispectral property of the multi-indexed
Laguerre and Jacobi orthogonal polynomials, \eqref{bispec}.

\subsection{Preparation}
\label{sec:prepare}

\subsubsection{some algebra}
\label{sec:alg}

Let us consider operators $A$, $B$ and $O_j$ ($j=1,2,\ldots$), which satisfy
\begin{align}
  &[A,B]=1+O_{1,1},\quad O_{1,1}\in\mathcal{O}\eqdef\mathbb{C}
  [O_1,O_2,\ldots],\n
  &AO_j,O_jA,BO_j,O_jB,O_jO_k\in\mathcal{O}.
  \label{ABO}
\end{align}
Any element $F$ of the ring $\mathbb{C}[A,B,O_1,O_2,\ldots]$
is written as a finite sum $F=\sum\limits_{i,j\geq 0}F_{i,j}B^jA^i+O_F$
($F_{i,j}\in\mathbb{C}$, $O_F\in\mathcal{O}$).
It is easy to show the following identity ($i,j\in\mathbb{Z}_{\geq 0}$)
by induction,
\begin{equation}
  A^iB^j=\sum_{r=0}^{\min(i,j)}a^{ij}_rB^{j-r}A^{i-r}+O_{i,j},\quad
  a^{ij}_r\eqdef r!\genfrac{(}{)}{0pt}{}{i}{r}\genfrac{(}{)}{0pt}{}{j}{r}
  =a^{ji}_r,\quad O_{i,j}\in\mathcal{O}.
  \label{AiBj=}
\end{equation}
The explicit form of $O_{i,j}$ can be obtained by the recurrence relations,
\begin{equation}
  O_{i+1,j}=\sum_{r=0}^{\min(i,j)}a^{i,j}_rO_{1,j-r}A^{i-r}+AO_{i,j},\quad
  O_{i,j+1}=\sum_{r=0}^{\min(i,j)}a^{i,j}_rB^{j-r}O_{i-r,1}+O_{i,j}B,
  \label{Oij}
\end{equation}
with $O_{i,0}=O_{0,j}=0$. 

The algebra of operators $\partial_{\eta}$ (derivative by $\eta$) and
$\eta$ (multiplication by $\eta$), $[\partial_{\eta},\eta]=1$, is a special
case of the above, namely $O_j=0$.
Eq.\,\eqref{AiBj=} with $i\leftrightarrow j$ becomes
\begin{equation}
  \partial_{\eta}^j\circ\eta^i
  =\sum_{r=0}^{\min(i,j)}a^{ij}_r\eta^{i-r}\partial_{\eta}^{j-r}.
  \label{etaidetaj}
\end{equation}

\subsubsection{shift operators}
\label{sec:op}

In the bispectral property \eqref{bispec}, $\Delta_{X,\mathcal{D}}$ is a
certain shift operator of $n$.
Usually a formal shift operator, e.g. $n\to n+1$, is used but here we realize
shift operators as differential operators acting on smooth functions of $n$. 
For a function $f(n)$, the exponential of $a\partial_n$ ($a$ : constant) acts
on $f(n)$ as a shift operator,
\begin{equation}
  e^{a\partial_n}f(n)=f(n+a),
\end{equation}
because 
\begin{equation*}
  e^{a\partial_n}f(n)=\sum_{k=0}^{\infty}\frac{a^k}{k!}\partial_n^k\,f(n)
  =\sum_{k=0}^{\infty}\frac{a^k}{k!}\frac{d^kf}{dn^k}(n)
  =f(n+a).
\end{equation*}
We regard a polynomial $P_n(\eta)$ as a sum $\sum\limits_{j=0}^na_j(n)\eta^j$
and treat $n$ (upper limit of the sum) as a continuous variable in the
following way: a sum $\sum\limits_{j=1}^nf(n,j)$ is understood as
\begin{equation}
  \sum_{j=1}^nf(n,j)=\int_{\frac12}^{n+\frac12}\!\!dx
  \sum_{j=-\infty}^{\infty}\delta(x-j)\cdot f(n,x),
  \label{sum}
\end{equation}
where $\delta(x)$ is the Dirac delta function
($\int_{\frac12}^{n+\frac12}$ is replaced by
$\int_{-\frac12}^{n+\frac12}$ for $\sum\limits_{j=0}^nf(n,j)$).
Of course, only an integer shift is allowed for the upper limit of the sum.
After all the calculations are done, we can evaluate various quantities at
$n=0,1,2,\ldots$ (and $j=n,n-1,\ldots$).

The exponential operator $e^{a\partial_n}$ is a shift operator.
If a constant $a$ is replaced by a function $g(n)$, the exponential
operator $e^{g(n)\partial_n}$ is no longer a shift operator, e.g.
$e^{an\partial_n}f(n)=f(e^an)$.
Let us define a `normal ordered' exponential operator
$:e^{g(n)\partial_n}\!:$ as
\begin{equation}
  :e^{g(n)\partial_n}\!:\,\eqdef
  \sum_{k=0}^{\infty}\frac{g(n)^k}{k!}\partial_n^k.
  \label{:e:}
\end{equation}
This acts on $f(n)$ as a shift operator,
\begin{equation}
  :e^{g(n)\partial_n}\!:f(n)=f\bigl(n+g(n)\bigr),
\end{equation}
because we have ($f(n)=\sum\limits_{l=0}^{\infty}\frac{f_l}{l!}n^l$),
\begin{align*}
  &\quad:e^{g(n)\partial_n}\!:f(n)
  =\sum_{k=0}^{\infty}\frac{g(n)^k}{k!}
  \partial_n^k\sum_{l=0}^{\infty}\frac{f_l}{l!}n^l
  =\sum_{k=0}^{\infty}\sum_{l=0}^{\infty}\frac{g(n)^k}{k!}
  \frac{f_l}{l!}\genfrac{(}{)}{0pt}{}{l}{k}k!\,n^{l-k}\n
  &=\sum_{l=0}^{\infty}\frac{f_l}{l!}\sum_{k=0}^l
  \genfrac{(}{)}{0pt}{}{l}{k}g(n)^kn^{l-k}
  =\sum_{l=0}^{\infty}\frac{f_l}{l!}\bigl(n+g(n)\bigr)^l
  =f\bigl(n+g(n)\bigr).
\end{align*}
For a constant $a$, we have $:e^{a\partial_n}\!:\,=e^{a\partial_n}$.
We remark that $:e^{-(n+a)\partial_n}\!:$ ($a$ : constant) maps a function
of $n$ to a constant, $:e^{-(n+a)\partial_n}\!:f(n)=f(-a)$.
The product of normal ordered exponential operators is again a normal
ordered exponential operator,
\begin{equation}
  :e^{g_1(n)\partial_n}\!:\,:e^{g_2(n)\partial_n}\!:
  \,=\,:e^{(g_1\star g_2)(n)\partial_n}\!:,\quad
  (g_1\star g_2)(n)\eqdef g_1(n)+g_2\bigl(n+g_1(n)\bigr),
  \label{:e::e:}
\end{equation}
and associative ($(g_1\star g_2)\star g_3=g_1\star(g_2\star g_3)$ is easily
shown).
Eq.\eqref{:e::e:} is shown by
\begin{align*}
  &\quad:e^{g_1(n)\partial_n}\!:\,:e^{g_2(n)\partial_n}\!:f(n)
  =\,:e^{g_1(n)\partial_n}\!:f\bigl(n+g_2(n)\bigr)
  =f\bigl(n+g_1(n)+g_2(n+g_1(n))\bigr)\n
  &=\,:e^{(g_1(n)+g_2(n+g_1(n)))\partial_n}\!:f(n).
\end{align*}
We give another proof:
\begin{align*}
  &\quad:e^{g_1(n)\partial_n}\!:\,:e^{g_2(n)\partial_n}\!:
  \,=\sum_{k=0}^{\infty}\frac{g_1(n)^k}{k!}\partial_n^k\circ
  \sum_{l=0}^{\infty}\frac{g_2(n)^l}{l!}\partial_n^l
  =\sum_{k=0}^{\infty}\sum_{l=0}^{\infty}\frac{g_1(n)^k}{k!\,l!}
  \sum_{r=0}^k\genfrac{(}{)}{0pt}{}{k}{r}\bigl(g_2(n)^l\bigr)^{(r)}
  \partial_n^{k-r}\partial_n^l\n
  &=\sum_{m=0}^{\infty}\sum_{k=0}^m\sum_{r=0}^k
  \frac{g_1(n)^k}{k!\,(m-k)!}\genfrac{(}{)}{0pt}{}{k}{r}
  \bigl(g_2(n)^{m-k}\bigr)^{(r)}\partial_n^{m-r}
  =\sum_{r=0}^{\infty}\sum_{m=r}^{\infty}\sum_{k=0}^m
  \frac{g_1(n)^k}{k!\,(m-k)!}\genfrac{(}{)}{0pt}{}{k}{r}
  \bigl(g_2(n)^{m-k}\bigr)^{(r)}\partial_n^{m-r}\n
  &=\sum_{r=0}^{\infty}\sum_{s=0}^{\infty}\sum_{t=0}^s
  \frac{g_1(n)^{r+t}}{(r+t)!\,(s-t)!}\genfrac{(}{)}{0pt}{}{r+t}{r}
  \bigl(g_2(n)^{s-t}\bigr)^{(r)}\partial_n^s
  =\sum_{s=0}^{\infty}\sum_{t=0}^s\sum_{r=0}^{\infty}
  \frac{g_1(n)^{r+t}}{r!\,s!}\genfrac{(}{)}{0pt}{}{s}{t}
  \bigl(g_2(n)^{s-t}\bigr)^{(r)}\partial_n^s\n
  &=\sum_{s=0}^{\infty}\frac{1}{s!}
  \sum_{t=0}^s\genfrac{(}{)}{0pt}{}{s}{t}g_1(n)^t
  \sum_{r=0}^{\infty}\frac{g_1(n)^r}{r!}\bigl(g_2(n)^{s-t}\bigr)^{(r)}
  \partial_n^s
  =\sum_{s=0}^{\infty}\frac{1}{s!}
  \sum_{t=0}^s\genfrac{(}{)}{0pt}{}{s}{t}g_1(n)^t
  \Bigl(g_2\bigl(n+g_1(n)\bigr)\Bigr)^{s-t}\partial_n^s\n
  &=\sum_{s=0}^{\infty}\frac{1}{s!}
  \Bigl(g_1(n)+g_2\bigl(n+g_1(n)\bigr)\Bigr)^s\partial_n^s
  =\,:e^{(g_1(n)+g_2(n+g_1(n)))\partial_n}\!:,
\end{align*}
where $(f(n))^{(r)}=\partial_n^rf(n)$.
Later we will use the following ($a,b$ : constants):
\begin{align}
  &:e^{a\partial_n}\!:\,:e^{b\partial_n}\!:\,=\,:e^{(a+b)\partial_n}\!:,\quad
  :e^{a\partial_n}\!:\,:e^{-(n+b)\partial_n}\!:\,=\,:e^{-(n+b)\partial_n}\!:,\n
  &:e^{-(n+b)\partial_n}\!:\,:e^{a\partial_n}\!:\,
  =\,:e^{-(n-a+b)\partial_n}\!:,\quad
  :e^{-(n+a)\partial_n}\!:\,:e^{-(n+b)\partial_n}\!:\,
  =\,:e^{-(n+b)\partial_n}\!:.
  \label{e:e:}
\end{align}

The product of the shift operator $e^{\pm k\partial_n}$
($k$: constant, integer) and a sum of functions $\sum\limits_{j=1}^nf(n,j)$
as an operator is
\begin{equation}
  e^{\pm k\partial_n}\circ\sum_{j=1}^nf(n,j)
  =\sum_{j=1}^{n\pm k}f(n\pm k,j)e^{\pm k\partial_n}.
\end{equation}
This is understood in the following way.
By rewriting the sum as \eqref{sum},
\begin{equation*}
  e^{\pm k\partial_n}\circ\sum_{j=1}^nf(n,j)
  =\int_{\frac12}^{n\pm k+\frac12}\!\!dx
  \sum_{j=-\infty}^{\infty}\delta(x-j)\cdot f(n\pm k,x)e^{\pm k\partial_n}
  =\sum_{j=1}^{n\pm k}f(n\pm k,j)e^{\pm k\partial_n}.
\end{equation*}

\subsection{Map $\flat$}
\label{sec:mapb}

By modifying the arguments in \cite{gkkm15}, we define the map $\flat$
\eqref{flat}.
In this subsection we consider arbitrary (ordinary) orthogonal polynomials
$P_n(\eta)$ in continuous variable $\eta$.
The polynomial $P_n(\eta)=c_n\eta^n+(\text{lower degree terms})$ is
defined by the three term recurrence relations \cite{szego}
\begin{equation}
  \eta P_n(\eta)=A_nP_{n+1}(\eta)+B_nP_n(\eta)+C_nP_{n-1}(\eta).
  \label{3trr}
\end{equation}
We set $P_0(\eta)\eqdef1$, $P_n(\eta)\eqdef0$ ($n<0$) and $A_{-1}\eqdef0$.
We assume that $A_n$, $B_n$ and $C_n$ are given as functions of continuous
$n$.
Note that \eqref{3trr} holds for $n\in\mathbb{Z}$ and $P_n(\eta)$ may not
satisfy any differential equation.

Since $\partial_{\eta}P_n(\eta)$ is a polynomial of degree $n-1$,
it can be written as
\begin{equation}
  \partial_{\eta}P_n(\eta)=\sum_{k=1}^nc_{n,k}P_{n-k}(\eta),
  \label{cnk}
\end{equation}
where $c_{n,k}$ are constants and we set $c_{n,0}\eqdef0$.
We have $A_n=\frac{c_n}{c_{n+1}}$ and $c_{n,1}=n\frac{c_n}{c_{n-1}}$,
which imply $A_nc_{n+1,1}=n+1$.
Let us define operators $\Delta$, $\Gamma$ and $O_j$ ($j=1,2,\ldots$) as
\begin{equation}
  \Delta\eqdef A_ne^{\partial_n}+B_n+C_ne^{-\partial_n},\quad
  \Gamma\eqdef\sum_{k=1}^nc_{n,k}:e^{-k\partial_n}\!:,\quad
  O_j\eqdef\,:e^{-(n+j)\partial_n}\!:.
  \label{Delta}
\end{equation}
We remark that $\Delta$, $\Gamma$ and $O_j$ commute with $\eta$ and
$\partial_{\eta}$.
{}From \eqref{3trr}, \eqref{cnk} and $P_n(\eta)=0$ ($n<0$),
they act on $P_n(\eta)$ as follows:
\begin{equation}
  \Delta P_n(\eta)=\eta P_n(\eta),\quad
  \Gamma P_n(\eta)=\partial_{\eta}P_n(\eta),\quad
  O_jP_n(\eta)=0.
  \label{Delta=eta}
\end{equation}
Let us calculate the commutation relation of $\Delta$ and $\Gamma$.
Results in \S\,\ref{sec:op} give
\begin{align}
  \Delta\Gamma&=A_n\sum_{k=1}^{n+1}c_{n+1,k}:e^{-(k-1)\partial_n}\!:
  +B_n\sum_{k=1}^nc_{n,k}:e^{-k\partial_n}\!:
  +C_n\sum_{k=1}^{n-1}c_{n-1,k}:e^{-(k+1)\partial_n}\!:\n
  &=A_nc_{n+1,1}+A_n\sum_{k=2}^{n+1}c_{n+1,k}:e^{-(k-1)\partial_n}\!:
  +B_n\sum_{k=1}^nc_{n,k}:e^{-k\partial_n}\!:
  +C_n\sum_{k=0}^{n-1}c_{n-1,k}:e^{-(k+1)\partial_n}\!:\n
  &=n+1+\sum_{k=1}^n(A_nc_{n+1,k+1}+B_nc_{n,k}+C_nc_{n-1,k-1})
  :e^{-k\partial_n}\!:,\\
  \Gamma\Delta&=\sum_{k=1}^nc_{n,k}A_{n-k}:e^{-(k-1)\partial_n}\!:
  +\sum_{k=1}^nc_{n,k}B_{n-k}:e^{-k\partial_n}\!:
  +\sum_{k=1}^nc_{n,k}C_{n-k}:e^{-(k+1)\partial_n}\!:\n
  &=A_{n-1}c_{n,1}+\sum_{k=2}^{n+1}A_{n-k}c_{n,k}:e^{-(k-1)\partial_n}\!:
  +\sum_{k=1}^nB_{n-k}c_{n,k}:e^{-k\partial_n}\!:\n
  &\quad
  +\sum_{k=0}^{n-1}C_{n-k}c_{n,k}:e^{-(k+1)\partial_n}\!:
  +C_0c_{n,n}:e^{-(n+1)\partial_n}\!:\n
  &=n+\sum_{k=1}^n(A_{n-k-1}c_{n,k+1}+B_{n-k}c_{n,k}+C_{n-k+1}c_{n,k-1})
  :e^{-k\partial_n}\!:
  +C_0c_{n,n}O_1,
\end{align}
where we have used $A_{-1}=0$, $c_{n,0}=0$ and $A_nc_{n+1,1}=n+1$.
{}From these we have
\begin{equation}
  [\Delta,\Gamma]=1+\sum_{k=1}^nb_{n,k}:e^{-k\partial_n}\!:-C_0c_{n,n}O_1,
  \label{[D,G]1}
\end{equation}
where the constant $b_{n,k}$ is
\begin{equation}
  b_{n,k}\eqdef A_nc_{n+1,k+1}-A_{n-k-1}c_{n,k+1}+(B_n-B_{n-k})c_{n,k}
  +C_nc_{n-1,k-1}-C_{n-k+1}c_{n,k-1}.
  \label{bnk}
\end{equation}
The l.h.s of \eqref{[D,G]1} acts on $P_n(\eta)$ as
\begin{align}
  &\quad(\text{l.h.s})P_n(\eta)=[\Delta,\Gamma]P_n(\eta)
  =(\Delta\Gamma-\Gamma\Delta)P_n(\eta)
  =(\Delta\partial_{\eta}-\Gamma\eta)P_n(\eta)\n
  &=(\partial_{\eta}\Delta-\eta\Gamma)P_n(\eta)
  =(\partial_{\eta}\eta-\eta\partial_{\eta})P_n(\eta)
  =[\partial_{\eta},\eta]P_n(\eta)
  =P_n(\eta),
\end{align}
and the r.h.s \eqref{[D,G]1} acts on $P_n(\eta)$ as
\begin{equation}
  (\text{r.h.s})P_n(\eta)
  =P_n(\eta)+\sum_{k=1}^nb_{n,k}P_{n-k}(\eta),
\end{equation}
which means $\sum\limits_{k=1}^nb_{n,k}P_{n-k}(\eta)=0$, namely $b_{n,k}=0$.
(This $b_{n,k}=0$ can be checked by explicit calculation of \eqref{bnk}.
We have checked this for $n\leq 15$ by using Mathematica.)
Therefore we obtain
\begin{equation}
  [\Delta,\Gamma]=1-C_0c_{n,n}O_1.
  \label{[D,G]}
\end{equation}
Products of $O_j$ and ($\Delta$, $\Gamma$, $O_k$) are
\begin{align}
  &\Delta O_j=(A_n+B_n+C_n)O_j,\quad
  O_j\Delta=A_{-j}O_{j-1}+B_{-j}O_j+C_{-j}O_{j+1},\\
  &\Gamma O_j=\Bigl(\sum_{k=1}^nc_{n,k}\Bigr)O_j,\quad
  O_j\Gamma_j=0,\\
  &O_jO_k=O_k,
  \label{OjOk}
\end{align}
and all of them belong to $\mathcal{O}=\mathbb{C}[O_1,O_2,\ldots]$
($O_1\Delta=B_{-1}O_1+C_{-1}O_2$ due to $A_{-1}=0$).
The relations \eqref{[D,G]}--\eqref{OjOk} satisfy the conditions given
in \S\,\ref{sec:alg}, where the correspondence is
$(A,B,O_j)\leftrightarrow(\Delta,\Gamma,O_j)$.
If $C_{-j}\neq 0$, we have
$\mathbb{C}[\Delta,\Gamma,O_1,O_2,\ldots]=\mathbb{C}[\Delta,\Gamma]$.

Any element $F$ of $\mathbb{C}[\partial_{\eta},\eta]$ is written as
a finite sum $F=\sum\limits_{i,j\geq 0}F_{i,j}\eta^i\partial_{\eta}^j$
($F_{i,j}\in\mathbb{C}$).
Let us define a map
$\flat:\mathbb{C}[\partial_{\eta},\eta]\to\mathbb{C}[\Delta,\Gamma]$:
\begin{equation}
  F=\sum_{i,j\geq 0}F_{i,j}\eta^i\partial_{\eta}^j
  \in\mathbb{C}[\partial_{\eta},\eta],\quad
  \flat(F)\eqdef\sum_{i,j\geq 0}F_{i,j}\Gamma^j\Delta^i
  \in\mathbb{C}[\Delta,\Gamma].
  \label{flat}
\end{equation}
We remark that $\flat(\partial_{\eta}\eta)$ is not directly given in the
above definition and it is calculated as
$\flat(\partial_{\eta}\eta)=\flat(\eta\partial_{\eta}+1)=\Gamma\Delta+1
\neq\Delta\Gamma=\Gamma\Delta+1-C_0c_{n,n}O_1$.
{}From \eqref{Delta=eta}, we have
\begin{equation}
  \eta^i\partial_{\eta}^jP_n(\eta)=\eta^i\Gamma^jP_n(\eta)
  =\Gamma^j\eta^iP_n(\eta)=\Gamma^j\Delta^iP_n(\eta).
\end{equation}
By using \eqref{etaidetaj} and \eqref{AiBj=} with $(A,B)=(\Delta,\Gamma)$,
we have
\begin{align}
  &\quad\eta^{i_2}\partial_{\eta}^{j_2}\eta^{i_1}
  \partial_{\eta}^{j_1}P_n(\eta)
  =\eta^{i_2}\Bigl(\sum_{r=0}^{\min(j_2,i_1)}a^{j_2i_1}_r
  \eta^{i_1-r}\partial_{\eta}^{j_2-r}\Bigr)\partial_{\eta}^{j_1}P_n(\eta)\n
  &=\sum_{r=0}^{\min(j_2,i_1)}a^{j_2i_1}_r
  \eta^{i_1+i_2-r}\partial_{\eta}^{j_1+j_2-r}P_n(\eta)
  =\sum_{r=0}^{\min(j_2,i_1)}a^{j_2i_1}_r
  \Gamma^{j_1+j_2-r}\Delta^{i_1+i_2-r}P_n(\eta)\n
  &=\Gamma^{j_1}\Bigl(\sum_{r=0}^{\min(i_1,j_2)}a^{i_1j_2}_r
  \Gamma^{j_2-r}\Delta^{i_1-r}\Bigr)\Delta^{i_2}P_n(\eta)
  =\Gamma^{j_1}(\Delta^{i_1}\Gamma^{j_2}-O_{i_1,j_2})\Delta^{i_2}P_n(\eta)\n
  &=\Gamma^{j_1}\Delta^{i_1}\Gamma^{j_2}\Delta^{i_2}P_n(\eta)
  -\Gamma^{j_1}O_{i_1,j_2}\Delta^{i_2}P_n(\eta)
  =\Gamma^{j_1}\Delta^{i_1}\Gamma^{j_2}\Delta^{i_2}P_n(\eta),
\end{align}
where we have used $O_{i_1,j_2}\Delta^{i_2}\in\mathcal{O}$ in the last line.
Therefore we obtain the following proposition.
\begin{prop}
The action of $\mathbb{C}[\partial_{\eta},\eta]$ on $P_n(\eta)$ is related to
that of $\mathbb{C}[\Delta,\Gamma]$ by the map $\flat$:
\begin{align}
  &F\in\mathbb{C}[\partial_{\eta},\eta]\Rightarrow
  FP_n(\eta)=\flat(F)P_n(\eta),
  \label{FPn}\\
  &F,G\in\mathbb{C}[\partial_{\eta},\eta]\Rightarrow
  FGP_n(\eta)=\flat(FG)P_n(\eta)=\flat(G)\flat(F)P_n(\eta).
  \label{FGPn}
\end{align}
  \label{propmapb}
\end{prop}
\vspace*{-10mm}
{\bf Remark 1}$\,$
This anti-homomorphism property \eqref{FGPn} does not hold as algebra,
namely $\flat(FG)\neq\flat(G)\flat(F)$ in general.\\
{\bf Remark 2}$\,$
The polynomial $P_n(\eta)$ may depend on a set of parameters
$\bm{\lambda}=(\lambda_1,\lambda_2,\ldots)$,
$P_n(\eta)=P_n(\eta;\bm{\lambda})$.
The above $\mathbb{C}[\partial_{\eta},\eta]$ and
$\mathbb{C}[\Delta,\Gamma]$ are understood as
$\mathbb{C}(\bm{\lambda})[\partial_{\eta},\eta]$ and
$\mathbb{C}(\bm{\lambda})[\Delta,\Gamma]$ respectively.

For later use, we present $\Delta^i$ and $\Gamma^j$ ($i,j=0,1,2,\ldots$):
\begin{align}
  \Delta^i&=\sum_{k=-i}^iD^{i,k}_ne^{k\partial_n},\quad
  D^{0,0}_n=1,\quad D^{i,k}_n\eqdef 0\ \ (|k|>i),\n
  &\quad D^{i,k}_n=D^{i-1,k-1}_nA_{n+k-1}+D^{i-1,k}_nB_{n+k}
  +D^{i-1,k+1}_nC_{n+k+1}\ \ (-i\leq k\leq i),
  \label{Deltai}\\
  \Gamma^j&=\sum_{k_1=1}^n\sum_{k_2=1}^{k_1-1}\,\sum_{k_3=1}^{k_1-k_2-1}
  \,\sum_{k_4=1}^{k_1-k_2-k_3-1}\cdots
  \!\!\sum_{k_j=1}^{k_1-k_2-\cdots-k_{j-1}-1}
  \!\!\!\!\!\!\!\!\!c_{n,k_2}c_{n-k_2,k_3}c_{n-k_2-k_3,k_4}\cdots
  c_{n-k_2-k_3-\cdots-k_{j-1},k_j}\n
  &\quad\times
  c_{n-k_2-k_3-\cdots-k_j,k_1-k_2-k_3-\cdots-k_j}
  :e^{-k_1\partial_n}\!:.
  \label{Gammaj}
\end{align}
Note that $\sum\limits_{k_1=1}^n$ is actually $\sum\limits_{k_1=j}^n$
because of our convention of the summation symbol:
$\sum\limits_{k=m}^{m-1}*=0$.
We explain $\Gamma^2$:
\begin{align}
  \Gamma^2&=\Bigl(\sum_{k_2=1}^nc_{n,k_2}:e^{-k_2\partial_n}\!:\Bigr)
  \Bigl(\sum_{k'_2=1}^nc_{n,k'_2}:e^{-k'_2\partial_n}\!:\Bigr)
  =\sum_{k_2=1}^nc_{n,k_2}\sum_{k'_2=1}^{n-k_2}c_{n-k_2,k'_2}
  :e^{-k_2\partial_n}\!:\,:e^{-k'_2\partial_n}\!:\n
  &=\sum_{k_2=1}^nc_{n,k_2}\sum_{k'_2=1}^{n-k_2}c_{n-k_2,k'_2}
  :e^{-(k_2+k'_2)\partial_n}\!:\,
  =\sum_{k_1=1}^n\sum_{k_2=1}^{k_1-1}c_{n,k_2}c_{n-k_2,k_1-k_2}
  :e^{-k_1\partial_n}\!:.
\end{align}
Here we have used $:e^{-k_2\partial_n}\!:\,:e^{-k'_2\partial_n}\!:\,
=\,:e^{-(k_2+k'_2)\partial_n}\!:$
because $k_2$ and $k'_2$ are independent of $n$.
As remarked in the first paragraph in \S\,\ref{sec:op},
after all the calculations are done, we can evaluate various quantities at
$k_2=n,n-1,\ldots$, $k'_2=n-k_2,\ldots$, etc.

\bigskip

In the rest of this subsection we present the explicit forms of $\Delta$ and
$\Gamma$ for the Hermite, Laguerre and Jacobi polynomials.

\subsubsection{example 1 : Hermite polynomial}
\label{sec:ex1:H}

The Hermite polynomial $H_n(\eta)$ \cite{szego} satisfies \eqref{3trr} with
\begin{equation}
  A_n=\tfrac12,\quad B_n=0,\quad C_n=n,
  \label{3trrH}
\end{equation}
and
\begin{equation}
  \partial_{\eta}H_n(\eta)=2nH_{n-1}(\eta).
\end{equation}
Therefore $\Delta$ and $\Gamma$ become
\begin{equation}
  \Delta=\tfrac12e^{\partial_n}+ne^{-\partial_n},\quad
  \Gamma=2ne^{-\partial_n},
  \label{DG:H}
\end{equation}
and they satisfy
\begin{equation}
  [\Delta,\Gamma]=1.
\end{equation}
In this case $\flat$ is an anti-isomorphism of algebra,
$\flat(FG)=\flat(G)\flat(F)$ \cite{gkkm15}.

\subsubsection{example 2 : Laguerre polynomial}
\label{sec:ex2:L}

The Laguerre polynomial $L^{(\alpha)}_n(\eta)$ \cite{szego} satisfies
\eqref{3trr} with
\begin{equation}
  A_n=-(n+1),\quad B_n=2n+\alpha+1,\quad C_n=-(n+\alpha),
  \label{3trrL}
\end{equation}
and
\begin{align}
  &\partial_{\eta}L^{(\alpha)}_n(\eta)=-L^{(\alpha+1)}_{n-1}(\eta),
  \label{dL}\\
  &L^{(\alpha)}_{n-1}(\eta)+L^{(\alpha-1)}_n(\eta)=L^{(\alpha)}_n(\eta).
  \label{idL}
\end{align}
{}From \eqref{idL} we have
\begin{equation}
  L^{(\alpha+1)}_n(\eta)=\sum_{k=0}^nL^{(\alpha)}_k(\eta).
\end{equation}
So we have $c_{n,k}=-1$ ($1\leq k\leq n$).
Therefore $\Delta$ and $\Gamma$ become
\begin{equation}
  \Delta=-(n+1)e^{\partial_n}+2n+\alpha+1-(n+\alpha)e^{-\partial_n},\quad
  \Gamma=-\sum_{k=1}^n:e^{-k\partial_n}\!:.
  \label{DG:L}
\end{equation}
It is easy to check that $b_{n,k}$ \eqref{bnk} vanishes.
The operators $\Delta$ and $\Gamma$ satisfy
\begin{equation}
  [\Delta,\Gamma]=1-\alpha O_1.
\end{equation}

\subsubsection{example 3 : Jacobi polynomial}
\label{sec:ex3:J}

The Jacobi polynomial $J^{(\alpha,\beta)}_n(\eta)$ \cite{szego} satisfies
\eqref{3trr} with
\begin{align}
  &A_n=\frac{2(n+1)(n+\alpha+\beta+1)}
  {(2n+\alpha+\beta+1)(2n+\alpha+\beta+2)},\quad
  B_n=\frac{\beta^2-\alpha^2}{(2n+\alpha+\beta)(2n+\alpha+\beta+2)},\n
  &C_n=\frac{2(n+\alpha)(n+\beta)}{(2n+\alpha+\beta)(2n+\alpha+\beta+1)},
  \label{3trrJ}
\end{align}
and
\begin{align}
  &\partial_{\eta}P^{(\alpha,\beta)}_n(\eta)
  =\tfrac12(n+\alpha+\beta+1)P^{(\alpha+1,\beta+1)}_{n-1}(\eta),
  \label{dJ}\\
  &(2n+\alpha+\beta)P^{(\alpha-1,\beta)}_n(\eta)
  =(n+\alpha+\beta)P^{(\alpha,\beta)}_n(\eta)
  -(n+\beta)P^{(\alpha,\beta)}_{n-1}(\eta),
  \label{idJ1}\\
  &(2n+\alpha+\beta)P^{(\alpha,\beta-1)}_n(\eta)
  =(n+\alpha+\beta)P^{(\alpha,\beta)}_n(\eta)
  +(n+\alpha)P^{(\alpha,\beta)}_{n-1}(\eta).
  \label{idJ2}
\end{align}
{}From \eqref{idJ1}--\eqref{idJ2} we have
\begin{equation}
  P^{(\alpha+1,\beta+1)}_n(\eta)
  =\alpha_nP^{(\alpha,\beta)}_n(\eta)
  +\beta_nP^{(\alpha+1,\beta+1)}_{n-1}(\eta)
  +\gamma_nP^{(\alpha+1,\beta+1)}_{n-2}(\eta)\quad(n\geq 0),
  \label{idJ3}
\end{equation}
where $\alpha_n$, $\beta_n$ and $\gamma_n$ are
\begin{align}
  &\alpha_n=\frac{(2n+\alpha+\beta+1)(2n+\alpha+\beta+2)}
  {(n+\alpha+\beta+1)(n+\alpha+\beta+2)},\quad
  \beta_n=\frac{(\beta-\alpha)(2n+\alpha+\beta+1)}
  {(n+\alpha+\beta+2)(2n+\alpha+\beta)},\n
  &\gamma_n=\frac{(n+\alpha)(n+\beta)(2n+\alpha+\beta+2)}
  {(n+\alpha+\beta+1)(n+\alpha+\beta+2)(2n+\alpha+\beta)}.
\end{align}
By substituting \eqref{idJ3} into the second term of the r.h.s of \eqref{idJ3}
and repeating this, $P^{(\alpha+1,\beta+1)}_n(\eta)$ has the following form
\begin{equation}
  P^{(\alpha+1,\beta+1)}_n(\eta)
  =p^{(k)}_n(\eta)+\beta^{(k)}_nP^{(\alpha+1,\beta+1)}_{n-k}(\eta)
  +\gamma^{(k)}_nP^{(\alpha+1,\beta+1)}_{n-k-1}(\eta),
\end{equation}
and we have
\begin{align*}
  &\quad P^{(\alpha+1,\beta+1)}_n(\eta)\\
  &=p^{(k)}_n(\eta)+\beta^{(k)}_n\bigl(
  \alpha_{n-k}P^{(\alpha,\beta)}_{n-k}(\eta)
  +\beta_{n-k}P^{(\alpha+1,\beta+1)}_{n-k-1}(\eta)
  +\gamma_{n-k}P^{(\alpha+1,\beta+1)}_{n-k-2}(\eta)\bigr)
  +\gamma^{(k)}_nP^{(\alpha+1,\beta+1)}_{n-k-1}(\eta)\\
  &=p^{(k)}_n(\eta)+\alpha_{n-k}\beta^{(k)}_nP^{(\alpha,\beta)}_{n-k}(\eta)
  +(\beta_{n-k}\beta^{(k)}_n+\gamma^{(k)}_n)
  P^{(\alpha+1,\beta+1)}_{n-k-1}(\eta)
  +\gamma_{n-k}\beta^{(k)}_nP^{(\alpha+1,\beta+1)}_{n-k-2}(\eta)\\
  &=p^{(k+1)}_n(\eta)+\beta^{(k+1)}_nP^{(\alpha+1,\beta+1)}_{n-k-1}(\eta)
  +\gamma^{(k+1)}_nP^{(\alpha+1,\beta+1)}_{n-k-2}(\eta).
\end{align*}
Namely $p^{(k)}_n(\eta)$, $\beta^{(k)}_n$ and $\gamma^{(k)}_n$ satisfy
the recurrence relations:
\begin{align}
  &p^{(k+1)}_n(\eta)=p^{(k)}_n(\eta)
  +\alpha_{n-k}\beta^{(k)}_nP^{(\alpha,\beta)}_{n-k}(\eta),\n
  &\beta^{(k+1)}_n=\beta_{n-k}\beta^{(k)}_n+\gamma^{(k)}_n,\quad
  \gamma^{(k+1)}_n=\gamma_{n-k}\beta^{(k)}_n\quad(1\leq k\leq n),
\end{align}
with the initial values,
\begin{equation}
  p^{(1)}_n(\eta)=\alpha_nP^{(\alpha,\beta)}_n(\eta),\quad
  \beta^{(1)}_n=\beta_n,\quad\gamma^{(1)}_n=\gamma_n.
\end{equation}
{}From this, $P^{(\alpha+1,\beta+1)}_n(\eta)=p^{(n+1)}_n(\eta)$ is
expressed as
\begin{equation}
  P^{(\alpha+1,\beta+1)}_n(\eta)=\sum_{k=0}^na^{(\alpha,\beta)}_{n,k}
  P^{(\alpha,\beta)}_{n-k}(\eta),\quad
  a^{(\alpha,\beta)}_{n,k}\eqdef\left\{
  \begin{array}{ll}
  \alpha_n&:k=0\\
  \alpha_{n-k}\beta^{(k)}_n&:1\leq k\leq n
  \end{array}\right.,
\end{equation}
and $c_{n,k}$ in \eqref{cnk} is given by
\begin{equation}
  c_{n,k}=\tfrac12(n+\alpha+\beta+1)a^{(\alpha,\beta)}_{n-1,k-1}.
\end{equation}
Therefore $\Delta$ and $\Gamma$ become
\begin{equation}
  \Delta=A_ne^{\partial_n}+B_n+C_ne^{-\partial_n},\quad
  \Gamma=\tfrac12(n+\alpha+\beta+1)\sum_{k=1}^na^{(\alpha,\beta)}_{n-1,k-1}
  :e^{-k\partial_n}\!:.
  \label{DG:J}
\end{equation}
We can check that $b_{n,k}$ \eqref{bnk} vanishes.
The operators $\Delta$ and $\Gamma$ satisfy
\begin{equation}
  [\Delta,\Gamma]=1-\tfrac12(n+\alpha+\beta+1)
  a^{(\alpha,\beta)}_{n-1,n-1}C_0O_1.
\end{equation}
Explicit forms of $a^{(\alpha,\beta)}_{n,k}$ for lower $k$ are
\begin{align}
  a^{(\alpha,\beta)}_{n,0}
  &=\frac{(2n+\alpha+\beta+1)_2}{(n+\alpha+\beta+1)_2},\quad
  a^{(\alpha,\beta)}_{n,1}
  =\frac{(\beta-\alpha)(2n+\alpha+\beta-1)(2n+\alpha+\beta+1)}
  {(n+\alpha+\beta)_3},\n
  a^{(\alpha,\beta)}_{n,2}
  &=\frac{(2n+\alpha+\beta-3)(2n+\alpha+\beta)
  \bigl((n+\alpha)(n+\beta)+(\alpha-\beta)^2-1\bigr)}
  {(n+\alpha+\beta-1)_4},\n
  a^{(\alpha,\beta)}_{n,3}
  &=\frac{(\beta-\alpha)(2n+\alpha+\beta-5)(2n+\alpha+\beta-1)}
  {(n+\alpha+\beta-2)_5}\n
  &\quad\times
  \bigl(2(n+\alpha+\beta)(n-1)+\alpha(\alpha+1)+\beta(\beta+1)-2\bigr),
  \label{ank01234}\\
  a^{(\alpha,\beta)}_{n,4}
  &=\frac{(2n+\alpha+\beta-7)(2n+\alpha+\beta-2)}{16(n+\alpha+\beta-3)_6}\n
  &\quad\times\Bigl(5(\alpha-\beta)^4
  +10(\alpha-\beta)^2\bigl(4n(n+\alpha+\beta-2)
  +(\alpha+\beta+1)(\alpha+\beta-5)+3\bigr)\n
  &\qquad
  +(2n+\alpha+\beta-6)(2n+\alpha+\beta-4)(2n+\alpha+\beta)(2n+\alpha+\beta+2)
  \Bigr).\nonumber
\end{align}

\subsection{Bispectral property}
\label{sec:bispec}

Following the arguments in \cite{gkkm15}, we discuss the bispectral property
of the multi-indexed Laguerre and Jacobi polynomials \eqref{bispec}.

{}From \eqref{FhDPn=}--\eqref{BhDPDn=}, the $M$-th order differential
operators $\hat{\mathcal{F}}^{(\mathcal{D})}$ and
$\hat{\mathcal{B}}^{(\mathcal{D})}$ are expressed as determinants:
\begin{equation}
  \hat{\mathcal{F}}^{(\mathcal{D})}
  =\rho^{(\mathcal{D})}_{\hat{\mathcal{F}}}(\eta)\left|
  \begin{array}{cccc}
  \mu_{d_1}&\cdots&\mu_{d_M}&1\\
  \mu_{d_1}^{(1)}&\cdots&\mu_{d_M}^{(1)}&\partial_{\eta}\\
  \vdots&\cdots&\vdots&\vdots\\
  \mu_{d_1}^{(M)}&\cdots&\mu_{d_M}^{(M)}&\partial_{\eta}^M
  \end{array}\right|,\quad
  \hat{\mathcal{B}}^{(\mathcal{D})}
  =\rho^{(\mathcal{D})}_{\hat{\mathcal{B}}}(\eta)\left|
  \begin{array}{cccc}
  m_1&\cdots&m_M&1\\
  m_1^{(1)}&\cdots&m_M^{(1)}&\partial_{\eta}\\
  \vdots&\cdots&\vdots&\vdots\\
  m_1^{(M)}&\cdots&m_M^{(M)}&\partial_{\eta}^M
  \end{array}\right|,
  \label{FhD}
\end{equation}
where $\mu_{\text{v}}^{(i)}=\partial_{\eta}^i\mu_{\text{v}}(\eta)$
and $m_j^{(i)}=\partial_{\eta}^im_j(\eta)$.
Our definition of the determinant (order of the matrix elements) is
\begin{equation}
  \det(a_{ij})_{1\leq i,j\leq n}=\left|
  \begin{array}{cccc}
  a_{11}&a_{12}&\cdots&a_{1n}\\
  \vdots&\vdots&\cdots&\vdots\\
  a_{n1}&a_{n2}&\cdots&a_{nn}
  \end{array}\right|
  =\sum_{i_1,\ldots,i_n=1}^n\!\!\varepsilon_{i_1i_2\ldots i_n}
  a_{i_11}a_{i_22}\cdots a_{i_nn},
\end{equation}
where $\varepsilon_{i_1i_2\ldots i_n}$ is the antisymmetric symbol.
{}From these forms and \eqref{FhDPn=}--\eqref{BhDPDn=}, the operator
$\hat{\mathcal{F}}^{(\mathcal{D})}$ belongs to
$\mathbb{C}[\partial_{\eta},\eta]$ but $\hat{\mathcal{B}}^{(\mathcal{D})}$
does not.
We have
\begin{equation}
  \hat{\mathcal{F}}^{(\mathcal{D})}P_n(\eta)=P_{\mathcal{D},n}(\eta),\quad
  \hat{\mathcal{B}}^{(\mathcal{D})}P_{\mathcal{D},n}(\eta)
  =\pi_{\mathcal{D}}(n)P_n(\eta),\quad
  \pi_{\mathcal{D}}(n)=\prod_{j=1}^M(\mathcal{E}_n-\tilde{\mathcal{E}}_{d_j}),
  \label{FhDPn}
\end{equation}
and these give the following:
\begin{equation}
  \hat{\mathcal{B}}^{(\mathcal{D})}\hat{\mathcal{F}}^{(\mathcal{D})}
  P_n(\eta)
  =\pi_{\mathcal{D}}(n)P_n(\eta),\quad
  \hat{\mathcal{F}}^{(\mathcal{D})}\hat{\mathcal{B}}^{(\mathcal{D})}
  P_{\mathcal{D},n}(\eta)
  =\pi_{\mathcal{D}}(n)P_{\mathcal{D},n}(\eta).
  \label{BhDFhDPn}
\end{equation}

For $X(\eta)$ \eqref{X=int}, the operator
$\Theta_{X,\mathcal{D}}=\hat{\mathcal{B}}^{(\mathcal{D})}\circ X(\eta)\circ
\hat{\mathcal{F}}^{(\mathcal{D})}$ belongs to
$\mathbb{C}[\partial_{\eta},\eta]$.
Therefore we can consider $\flat(\Theta_{X,\mathcal{D}})$.
Following the argument in \cite{gkkm15}, let us define
$\Delta_{X,\mathcal{D}}$,
\begin{equation}
  \Delta_{X,\mathcal{D}}\eqdef
  \flat(\Theta_{X,\mathcal{D}})\circ\pi^{-1}_{\mathcal{D}}(n),
  \label{DeltaXD}
\end{equation}
which commutes with $\eta$ and $\partial_{\eta}$.
Then we have a theorem.
\begin{thm}
For the multi-indexed Laguerre and Jacobi polynomials
$P_{\mathcal{D},n}(\eta)$ and a polynomial $X(\eta)$ \eqref{X=int}, we have
\begin{equation}
  X(\eta)P_{\mathcal{D},n}(\eta)
  =\Delta_{X,\mathcal{D}}P_{\mathcal{D},n}(\eta).
  \label{XP=DP}
\end{equation}
\label{thmbispec}
\end{thm}
\vspace*{-7mm}
\underline{Proof}$\,$
We have
\begin{align}
  &\quad\bigl(\hat{\mathcal{B}}^{(\mathcal{D})}X(\eta)\bigr)
  P_{\mathcal{D},n}(\eta)
  =\hat{\mathcal{B}}^{(\mathcal{D})}X(\eta)
  \hat{\mathcal{F}}^{(\mathcal{D})}P_n(\eta)
  =\bigl(\hat{\mathcal{B}}^{(\mathcal{D})}\circ X(\eta)\circ
  \hat{\mathcal{F}}^{(\mathcal{D})}\bigr)P_n(\eta)\n
  &=\flat\bigl(\hat{\mathcal{B}}^{(\mathcal{D})}\circ X(\eta)\circ
  \hat{\mathcal{F}}^{(\mathcal{D})}\bigr)P_n(\eta)
  =\bigl(\Delta_{X,\mathcal{D}}\circ\pi_{\mathcal{D}}(n)\bigr)P_n(\eta)
  =\Delta_{X,\mathcal{D}}\pi_{\mathcal{D}}(n)P_n(\eta)\n
  &=\Delta_{X,\mathcal{D}}\hat{\mathcal{B}}^{(\mathcal{D})}
  \hat{\mathcal{F}}^{(\mathcal{D})}P_n(\eta)
  =\bigl(\Delta_{X,\mathcal{D}}\hat{\mathcal{B}}^{(\mathcal{D})}\bigr)
  \hat{\mathcal{F}}^{(\mathcal{D})}P_n(\eta)
  =\bigl(\hat{\mathcal{B}}^{(\mathcal{D})}\Delta_{X,\mathcal{D}}\bigr)
  P_{\mathcal{D},n}(\eta),
\end{align}
where we have used \eqref{FhDPn}--\eqref{BhDFhDPn}, \eqref{FPn} and
$[\Delta_{X,\mathcal{D}},\hat{\mathcal{B}}^{(\mathcal{D})}]=0$.
Therefore we obtain
\begin{equation}
  \hat{\mathcal{B}}^{(\mathcal{D})}\bigl(X(\eta)
  -\Delta_{X,\mathcal{D}}\bigr)P_{\mathcal{D},n}(\eta)=0.
  \label{B(X-D)P=0}
\end{equation}
For appropriate parameter range, various operators appearing in each step
of the Darboux transformations are non-singular and we can use properties
of the inner product $(f,g)=\int_{x_1}^{x_2}dxf(x)g(x)$.
For any polynomial $\mathcal{P}(\eta)$ in $\eta$, we have
\begin{align}
  &\quad\bigl(\phi_{\mathcal{D}\,n}(x),
  \Psi_{\mathcal{D}}(x)\mathcal{P}(\eta(x))\bigr)
  =(\hat{\mathcal{A}}^{(\mathcal{D})}\phi_0(x)P_n(\eta(x)),
  \Psi_{\mathcal{D}}(x)\mathcal{P}(\eta(x))\bigr)
  =(\phi_0P_n,\hat{\mathcal{A}}^{(\mathcal{D})\,\dagger}
  \Psi_{\mathcal{D}}\mathcal{P}\bigr)\n
  &=(\phi_0P_n,\phi_0\hat{\mathcal{B}}^{(\mathcal{D})}\mathcal{P}\bigr)
  =(\phi_0^2P_n,\hat{\mathcal{B}}^{(\mathcal{D})}\mathcal{P}\bigr),
\end{align}
where we have used \eqref{Ahsphin}, \eqref{Bh(s)}, etc.
If $\hat{\mathcal{B}}^{(\mathcal{D})}\mathcal{P}=0$, we have
$(\phi_{\mathcal{D}\,n},\Psi_{\mathcal{D}}\mathcal{P})=0$ and the
completeness of $\phi_{\mathcal{D}\,n}$ implies $\mathcal{P}=0$.
We remark that this result is derived for appropriate parameter range
but it is valid for any parameter range because it is a relation of
a polynomial.
Therefore \eqref{B(X-D)P=0} gives \eqref{XP=DP}.
\hfill\fbox{}\medskip\\
{\bf Remark 1}$\,$
We have used \eqref{FPn} but not used \eqref{FGPn}.
For $\flat\bigl(\hat{\mathcal{B}}^{(\mathcal{D})}\circ X(\eta)\circ
\hat{\mathcal{F}}^{(\mathcal{D})}\bigr)$, we can not apply \eqref{FGPn},
because $\hat{\mathcal{B}}^{(\mathcal{D})}$ does not belong to
$\mathbb{C}[\partial_{\eta},\eta]$.
The commutativity
$[\Delta_{X,\mathcal{D}},\hat{\mathcal{B}}^{(\mathcal{D})}]=0$ is important.\\
{\bf Remark 2}$\,$
If we already know the coefficients $r_{n,k}^{X,\mathcal{D}}$
\eqref{XP}--\eqref{rnk0}, the operator $\Delta_{X,\mathcal{D}}$ is
expressed as
\begin{equation}
  \Delta_{X,\mathcal{D}}
  =\sum_{k=-L}^Lr_{n,k}^{X,\mathcal{D}}e^{k\partial_n}
  =\sum_{k=-L}^Lr_{n,k}^{(0)\,X,\mathcal{D}}e^{k\partial_n}\circ
  \pi_{\mathcal{D}}^{-1}(n).
\end{equation}

\subsection{Examples}
\label{sec:ex}

As an illustration of Theorem\,\ref{thmbispec}, we present examples:
$M=1$ case, $\mathcal{D}=\{d_1\}$.

Eqs.\eqref{FhD} give
\begin{equation}
  \hat{\mathcal{F}}^{(\mathcal{D})}
  =\rho^{(\mathcal{D})}_{\hat{\mathcal{F}}}\mu_{d_1}^2
  \partial_{\eta}\circ\mu_{d_1}^{-1},\quad
  \hat{\mathcal{B}}^{(\mathcal{D})}
  =\rho^{(\mathcal{D})}_{\hat{\mathcal{B}}}m_1^2
  \partial_{\eta}\circ m_1^{-1},
\end{equation}
and $\Theta_{X,\mathcal{D}}$ becomes
\begin{align}
  &\quad\Theta_{X,\mathcal{D}}
  =\rho^{(\mathcal{D})}_{\hat{\mathcal{B}}}m_1^2
  \partial_{\eta}\circ m_1^{-1}X
  \rho^{(\mathcal{D})}_{\hat{\mathcal{F}}}\mu_{d_1}^2
  \partial_{\eta}\circ\mu_{d_1}^{-1}\n
  &=\rho^{(\mathcal{D})}_{\hat{\mathcal{B}}}
  \rho^{(\mathcal{D})}_{\hat{\mathcal{F}}}Xm_1\mu_{d_1}\partial_{\eta}^2
  +\rho^{(\mathcal{D})}_{\hat{\mathcal{B}}}m_1^2\mu_{d_1}
  \partial_{\eta}\bigl(m_1^{-1}X\rho^{(\mathcal{D})}_{\hat{\mathcal{F}}}\bigr)
  \partial_{\eta}
  -\rho^{(\mathcal{D})}_{\hat{\mathcal{B}}}m_1^2
  \partial_{\eta}\bigl(m_1^{-1}X\rho^{(\mathcal{D})}_{\hat{\mathcal{F}}}
  \partial_{\eta}\mu_{d_1}\bigr).
\end{align}

\subsubsection{Laguerre}
\label{sec:exL}

Let us consider type $\I$ Laguerre case, $\mathcal{D}=\{d^{\I}_1\}$.
Then we have
$\Xi_{\mathcal{D}}(\eta)=L^{(g-\frac12)}_{d_1}(-\eta)\eqdef\xi(\eta)$ and
\begin{equation}
  \rho^{(\mathcal{D})}_{\hat{\mathcal{F}}}=e^{-\eta},
  \ \ \rho^{(\mathcal{D})}_{\hat{\mathcal{B}}}
  =-4\eta^{g+\frac32}\xi^{-1},
  \ \ \mu_{d_1}=e^{\eta}\xi,\ \ m_1=\eta^{-g-\frac12},
\end{equation}
and $\Theta_{X,\mathcal{D}}$ becomes
\begin{equation}
  -\tfrac14\Theta_{X,\mathcal{D}}
  =\eta X\partial_{\eta}^2
  +\bigl((g+\tfrac12-\eta)X+\eta\xi Y\bigr)\partial_{\eta}
  -(d_1+g+\tfrac12)X-\eta(\xi+\partial_{\eta}\xi)Y,
\end{equation}
where we have used $\partial_{\eta}X=\Xi_{\mathcal{D}}Y$ and
$\eta\partial_{\eta}^2\xi+(g+\frac12+\eta)\partial_{\eta}\xi=d_1\xi$.
For simplicity we take $d_1=1$ and a minimal degree one $X_{\text{min}}$,
which corresponds to $Y(\eta)=1$.
Then we have
\begin{align}
  X(\eta)&=X_{\text{min}}(\eta)=\tfrac12\eta(\eta+2g+1)
  =L^{(g-\frac32)}_2(-\eta)-L^{(g-\frac32)}_2(0),\\
  \Theta_{X,\mathcal{D}}
  &=\bigl(-2\eta^3-4(g+\tfrac12)\eta^2\bigr)\partial_{\eta}^2
  +\bigl(2\eta^3+2(g-\tfrac32)\eta^2-4(g+\tfrac12)_2\eta\bigr)
  \partial_{\eta}\n
  &\quad
  +2(g+\tfrac72)\eta^2+4(g+\tfrac32)^2\eta.
\end{align}
By the map $\flat$, $\Theta_{X,\mathcal{D}}$ is mapped to
\begin{align}
  \flat(\Theta_{X,\mathcal{D}})
  &=\Gamma^2\bigl(-2\Delta^3-4(g+\tfrac12)\Delta^2\bigr)
  +\Gamma\bigl(2\Delta^3+2(g-\tfrac32)\Delta^2-4(g+\tfrac12)_2\Delta\bigr)\n
  &\quad
  +2(g+\tfrac72)\Delta^2+4(g+\tfrac32)^2\Delta.
\end{align}
The operators $\Delta$ and $\Gamma$ are given in \eqref{DG:L} and
$\Gamma^2$ \eqref{Gammaj} is
$\Gamma^2=\sum\limits_{k=2}^n(k-1):e^{-k\partial_n}\!:$.
A straightforward calculation gives
\begin{align}
  \flat(\Theta_{X,\mathcal{D}})
  &=\tfrac12(n+2)_2\times4(n+g+\tfrac72)e^{2\partial_n}
  -(n+1)(2g+2n+3)\times4(n+g+\tfrac52)e^{\partial_n}\n
  &\quad
  +\tfrac18\bigl(24n^2+4(10g+11)n+(2g+1)(6g+13)\bigr)\times4(n+g+\tfrac32)\n
  &\quad
  -\tfrac12(2g+2n-1)(2g+2n+3)\times4(n+g+\tfrac12)e^{-\partial_n}\n
  &\quad
  +\tfrac18(2g+2n-3)(2g+2n+3)\times4(n+g-\tfrac12)e^{-2\partial_n}+O,
\end{align}
where $O$ is an element of $\mathbb{C}[O_1,O_2,\ldots]$
\begin{equation}
  O=(g-\tfrac12)^2\bigl(3(2g-3)n-8\bigr)O_1
  -4(g-\tfrac12)_2\bigl((2g-1)n-1\bigr)O_2
  +2n(g-\tfrac52)_3O_3,
\end{equation}
which annihilates $P_{\mathcal{D},n}(\eta)$.
By using \eqref{XP=DP} and $\pi_{\mathcal{D}}(n)=4(n+g+\tfrac32)$, we obtain
\begin{align}
  r_{n,2}^{X,\mathcal{D}}&=\tfrac12(n+1)_2,\quad
  r_{n,1}^{X,\mathcal{D}}=-(n+1)(2g+2n+3),\n
  r_{n,0}^{X,\mathcal{D}}
  &=\tfrac18\bigl(24n^2+4(10g+11)n+(2g+1)(6g+13)\bigr),
  \label{Ex1_L}\\
  r_{n,-1}^{X,\mathcal{D}}&=-\tfrac12(2g+2n-1)(2g+2n+3),\quad
  r_{n,-2}^{X,\mathcal{D}}=\tfrac18(2g+2n-3)(2g+2n+3).
  \nonumber
\end{align}
These 5-term recurrence relations were given in \cite{d14,mt14,rrmiop2}.

\subsubsection{Jacobi}
\label{sec:exJ}

Let us consider type $\I$ Jacobi case, $\mathcal{D}=\{d^{\I}_1\}$.
We set $a=g+h$ and $b=g-h$.
Then we have
$\Xi_{\mathcal{D}}(\eta)=P^{(g-\frac12,\frac12-h)}_{d_1}(\eta)
\eqdef\xi(\eta)$ and
\begin{equation}
  \rho^{(\mathcal{D})}_{\hat{\mathcal{F}}}
  =\bigl(\tfrac{1+\eta}{2}\bigr)^{h+\frac12},
  \ \ \rho^{(\mathcal{D})}_{\hat{\mathcal{B}}}
  =-16\bigl(\tfrac{1-\eta}{2}\bigr)^{g+\frac32}\xi^{-1},
  \ \ \mu_{d_1}=\bigl(\tfrac{1+\eta}{2}\bigr)^{\frac12-h}\xi,
  \ \ m_1=\bigl(\tfrac{1-\eta}{2}\bigr)^{-g-\frac12},
\end{equation}
and $\Theta_{X,\mathcal{D}}$ becomes
\begin{align}
  -\tfrac14\Theta_{X,\mathcal{D}}
  &=(1-\eta^2)X\partial_{\eta}^2
  +\bigl(-(b+(a+1)\eta)X+(1-\eta^2)\xi Y\bigr)\partial_{\eta}\\
  &\quad
  +\bigl(d_1(d_1+1+b)-(g+\tfrac12)(h-\tfrac12)\bigr)X
  +\bigl((h-\tfrac12)(1-\eta)\xi-(1-\eta^2)\partial_{\eta}\xi\bigr)Y,
  \nonumber
\end{align}
where we have used $\partial_{\eta}X=\Xi_{\mathcal{D}}Y$ and
$(1-\eta^2)\partial_{\eta}^2\xi
+\bigl(1-a-(b+2)\eta\bigr)\partial_{\eta}\xi
=-d_1(d_1+1+b)\xi$.
For simplicity we take $d_1=1$ and a minimal degree one $X_{\text{min}}$,
which corresponds to $Y(\eta)=1$.
Then we have
\begin{align}
  X(\eta)&=X_{\text{min}}(\eta)=
  \tfrac14\eta\bigl((b+2)\eta+2(a-1)\bigr)
  =\tfrac{2}{b+1}\bigl(P^{(g-\frac32,-h-\frac12)}_2(\eta)
  -P^{(g-\frac32,-h-\frac12)}_2(0)\bigr),\\
  \Theta_{X,\mathcal{D}}
  &=\bigl((b+2)\eta^4+2(a-1)\eta^3-(b+2)\eta^2-2(a-1)\eta\bigr)
  \partial_{\eta}^2\n
  &\quad
  +\bigl((a+3)(b+2)\eta^3+(2a^2+b^2+4g-4)\eta^2
  +2((a-2)b-2)\eta-2a+2\bigr)
  \partial_{\eta}\n
  &\quad
  +\tfrac14(2h-3)\bigl((b+2)(2g+7)\eta^2
  +2\bigl(a(a+b)+g+7h-5)\eta-4a-4\bigr).
\end{align}
By the map $\flat$, $\Theta_{X,\mathcal{D}}$ is mapped to
\begin{align}
  \flat(\Theta_{X,\mathcal{D}})
  &=\Gamma^2\bigl((b+2)\Delta^4+2(a-1)\Delta^3-(b+2)\Delta^2
  -2(a-1)\Delta\bigr)\n
  &\quad
  +\Gamma\bigl((a+3)(b+2)\Delta^3+(2a^2+b^2+4g-4)\Delta^2
  +2((a-2)b-2)\Delta-2a+2\bigr)\n
  &\quad
  +\tfrac14(2h-3)\bigl((b+2)(2g+7)\Delta^2
  +2\bigl(a(a+b)+4a-3b-5)\Delta-4a-4\bigr).
\end{align}
The operators $\Delta$ and $\Gamma$ are given in \eqref{DG:J} and
$\Delta^i$ and $\Gamma^2$ are given in \eqref{Deltai}--\eqref{Gammaj}.
By using \eqref{ank01234}, a straightforward but a little lengthy
calculation gives
\begin{align}
  &\quad\flat(\Theta_{X,\mathcal{D}})\n
  &=\frac{(n+1)_2(b+2)(a+n)_2(2h+2n-3)}{(a+2n)_4(2h+2n+1)}
  \times(2n+2g+7)(2n+2h+1)e^{2\partial_n}\n
  &\quad 
  +\frac{(n+1)(a-1)(a+n)(2g+2n+3)(2h+2n-3)}{(a+2n-1)_3(a+2n+3)}
  \times(2n+2g+5)(2n+2h-1)e^{\partial_n}\n
  &\quad
  +\frac{b+2}{4(a+2n-2)_2(a+2n+1)_2}\Bigl(
  -b(b+4)\bigl(2n(a+n)-(a-2)(a-1)\bigr)\n
  &\qquad
  +(a+2n-1)(a+2n+1)\bigl(2n(a+n)-(a-2)(2a-1)\bigr)\Bigr)
  \times(2n+2g+3)(2n+2h-3)\n
  &\quad
  +\frac{(a-1)(2g+2n-1)(2g+2n+3)(h+n-\tfrac32)_2}{(a+2n-3)(a+2n-1)_3}
  \times(2n+2g+1)(2n+2h-5)e^{-\partial_n}\n
  &\quad
  +\frac{(b+2)(2g+2n-3)(2g+2n+3)(h+n-\tfrac32)_2}{4(a+2n-3)_4}
  \times(2n+2g-1)(2n+2h-7)e^{-2\partial_n}\n
  &\quad
  +\sum_{k=3}^n(\cdots):e^{-k\partial_n}\!:+O,
\end{align}
where $O$ is an element of $\mathbb{C}[O_1,O_2,\ldots]$.
{}From Theorem\,\ref{conj_oQM} and \ref{thmbispec}, the coefficients
$(\cdots)$ in the sum $\sum_{k=3}^n$ should vanish.
By using \eqref{XP=DP} and $\pi_{\mathcal{D}}(n)=(2n+2g+3)(2n+2h-3)$,
we obtain
\begin{align}
  r_{n,2}^{X,\mathcal{D}}&=
  \frac{(n+1)_2(b+2)(a+n)_2(2h+2n-3)}{(a+2n)_4(2h+2n+1)},\n
  r_{n,1}^{X,\mathcal{D}}&=
  \frac{(n+1)(a-1)(a+n)(2g+2n+3)(2h+2n-3)}{(a+2n-1)_3(a+2n+3)},\n
  r_{n,0}^{X,\mathcal{D}}&=
  \frac{b+2}{4(a+2n-2)_2(a+2n+1)_2}\Bigl(
  -b(b+4)\bigl(2n(a+n)-(a-2)(a-1)\bigr)\n
  &\qquad\qquad
  +(a+2n-1)(a+2n+1)\bigl(2n(a+n)-(a-2)(2a-1)\bigr)\Bigr),
  \label{Ex1_J}\\
  r_{n,-1}^{X,\mathcal{D}}&=
  \frac{(a-1)(2g+2n-1)(2g+2n+3)(h+n-\tfrac32)_2}{(a+2n-3)(a+2n-1)_3},\n
  r_{n,-2}^{X,\mathcal{D}}&=
  \frac{(b+2)(2g+2n-3)(2g+2n+3)(h+n-\tfrac32)_2}{4(a+2n-3)_4},
  \nonumber
\end{align}
which were given in \cite{rrmiop2} ($g=h$ case was given in \cite{mt14}).

\section{Summary and Comments}
\label{sec:summary}

The recurrence relations with constant coefficients for the multi-indexed
Laguerre and Jacobi orthogonal polynomials conjectured in our previous
paper $\II$ \cite{rrmiop2} are established as Theorem\,\ref{conj_oQM} by
following the argument in \cite{gkkm15}.
Their bispectral properties are also discussed by the similar argument in
\cite{gkkm15} and Theorem\,\ref{thmbispec} is obtained.
To obtain this, the map $\flat$ plays an important role but it is not an
anti-isomorphism in contrast to the exceptional Hermite case in \cite{gkkm15}.
The discussion in \S\,\ref{sec:mapb} is valid for any ordinary orthogonal
polynomials.

{}From Theorem\,\ref{thmbispec}, we can obtain the coefficients
$r_{n,k}^{X,\mathcal{D}}$ explicitly as demonstrated in \S\,\ref{sec:ex},
because $\hat{\mathcal{F}}^{(\mathcal{D})}$,
$\hat{\mathcal{B}}^{(\mathcal{D})}$, $X(\eta)$, $\Xi(\eta)$, $\Delta$,
$\Gamma$ and $\pi_{\mathcal{D}}(n)$ are known as \eqref{FhD}, \eqref{X=int},
\eqref{Xis}, \eqref{DG:L}, \eqref{DG:J} and \eqref{piD}.
In practice, however, this calculation is not so easy.
The examples in \cite{rrmiop2} were obtained by a brute force method:
Expand $X(\eta)P_{\mathcal{D},n}(\eta)$ in terms of $P_{\mathcal{D},m}(\eta)$
for small $n$, and guess $r_{n,k}^{X,\mathcal{D}}$ for arbitrary $n$
(Or, based on \eqref{rnk0}, calculate $\Theta_{X,\mathcal{D}}P_n(\eta)$ and
expand it in terms of $P_m(\eta)$ for small $n$, and guess
$r_{n,k}^{(0)\,X,\mathcal{D}}$ for arbitrary $n$).
We hope to find a more efficient method to obtain $r_{n,k}^{X,\mathcal{D}}$.

In our previous paper $\II$ \cite{rrmiop2}, the recurrence relations with
constant coefficients are conjectured also for the multi-indexed Wilson
and Askey-Wilson orthogonal polynomials.
These polynomials satisfy second order difference equations.
The method in the present paper may be applied to these polynomials but
it seems more difficult technically.
We hope this problem will be solved in the near future.

\section*{Acknowledgments}

I thank R.\,Sasaki for discussion and reading of the manuscript.
I am supported in part by Grant-in-Aid for Scientific Research
from the Ministry of Education, Culture, Sports, Science and Technology
(MEXT), No.25400395.

\bigskip
\appendix
\section{Darboux Transformation}
\label{app:DT}

In this appendix we review the algebraic aspects of the Darboux transformation
\cite{darb}.
We do not discuss non-singularity of operators, square integrability of
wavefunctions, etc.

Various formulas in the Darboux transformation are expressed in terms of the
Wronskian.
The Wronskian of a set of $n$ functions $\{f_j(x)\}$ is defined by
\begin{equation}
  \text{W}\,[f_1,\ldots,f_n](x)
  \eqdef\det\Bigl(\frac{d^{j-1}f_k(x)}{dx^{j-1}}\Bigr)_{1\leq j,k\leq n},
\end{equation}
(for $n=0$, we set $\text{W}\,[\cdot](x)=1$).
It satisfies the following identities ($n\geq 0$),
\begin{align}
  &\text{W}[gf_1,gf_2,\ldots,gf_n](x)
  =g(x)^n\text{W}[f_1,f_2,\ldots,f_n](x),
  \label{Wformula1}\\
  &\text{W}\bigl[\text{W}[f_1,f_2,\ldots,f_n,g],
  \text{W}[f_1,f_2,\ldots,f_n,h]\,\bigr](x)\n
  &\ \ =\text{W}[f_1,f_2,\ldots,f_n](x)\,
  \text{W}[f_1,f_2,\ldots,f_n,g,h](x),
  \label{Wformula2}\\
  &\text{W}[f_1,f_2,\ldots,f_n](x)
  =\Bigl(\frac{d\eta(x)}{dx}\Bigr)^{\frac12n(n-1)}
  \text{W}[F_1,F_2,\ldots,F_n]\bigl(\eta(x)\bigr),\n
  &\qquad\qquad\qquad\qquad\text{where }
  f_j(x)=F_j\bigl(\eta(x)\bigr),
  \label{Wformula3}\\
  &\text{W}[F_1,F_2,\ldots,F_n](x)
  =(-1)^{\frac12n(n-1)}\text{W}[f_1,f_2,\ldots,f_n](x)^{n-1},\n
  &\qquad\qquad\qquad\qquad\text{where }
  F_j(x)=\text{W}[f_1,\ldots,f_{j-1},f_{j+1},\ldots,f_n](x).
  \label{Wformula4}
\end{align}
We learned \eqref{Wformula4} in Ref.\cite{gkkm15}.

\subsection{Darboux transformation}
\label{app:DT_1step}

We consider the Schr\"odinger equation,
\begin{equation}
  \mathcal{H}\psi(x)=\mathcal{E}\psi(x),\quad
  \mathcal{H}=p^2+U(x),\quad p=-i\frac{d}{dx},\quad x_1<x<x_2.
\end{equation}
By taking a seed solution $\tilde{\phi}(x)$, which is any solution of the
Schr\"odinger equation,
\begin{equation}
  \mathcal{H}\tilde{\phi}(x)=\tilde{\mathcal{E}}\tilde{\phi}(x),
\end{equation}
the Hamiltonian $\mathcal{H}$ is expressed as
\begin{align}
  \mathcal{H}=\hat{\mathcal{A}}^{\dagger}\hat{\mathcal{A}}
  +\tilde{\mathcal{E}},\quad
  \hat{\mathcal{A}}\eqdef\tfrac{d}{dx}
  -\partial_x\log\bigl|\tilde{\phi}(x)\bigr|,\quad
  \hat{\mathcal{A}}^{\dagger}=-\bigl(
  \tfrac{d}{dx}-\partial_x\log\bigl|\tilde{\phi}^{-1}(x)\bigr|\bigr),
\end{align}
where $f^{-1}(x)$ means $f^{-1}(x)=f(x)^{-1}$.
We do not discuss the non-singularity of the operators $\hat{\mathcal{A}}$
and $\hat{\mathcal{A}}^{\dagger}$ as mentioned above.
The Darboux transformation is given by
\begin{equation}
  \mathcal{H}^{\text{new}}\eqdef\hat{\mathcal{A}}\hat{\mathcal{A}}^{\dagger}
  +\tilde{\mathcal{E}},\quad
  \psi^{\text{new}}(x)\eqdef\hat{\mathcal{A}}\psi(x).
  \label{DT}
\end{equation}
Then we have
\begin{align}
  &\mathcal{H}^{\text{new}}\psi^{\text{new}}(x)
  =\mathcal{E}\psi^{\text{new}}(x),\\
  &\mathcal{H}^{\text{new}}\tilde{\phi}^{-1}(x)
  =\tilde{\mathcal{E}}\tilde{\phi}^{-1}(x)\quad
  \bigl(\Leftarrow\hat{\mathcal{A}}^{\dagger}\tilde{\phi}^{-1}(x)=0\bigr),\\
  &\hat{\mathcal{A}}^{\dagger}\psi^{\text{new}}(x)
  =(\mathcal{E}-\tilde{\mathcal{E}})\psi(x).
  \label{Ahdpsinew}
\end{align}
The first and second equations say that $\psi^{\text{new}}$ and
$\tilde{\phi}^{-1}$ are solutions of the new Schr\"odinger equation, but
it does not mean that they exhaust all of the solutions.
We remark that the new wavefunction corresponding to the seed solution
is absent in the new system, because
$\tilde{\phi}^{\text{new}}(x)=\hat{\mathcal{A}}\tilde{\phi}(x)=0$.
The second equation of \eqref{DT} and \eqref{Ahdpsinew} are expressed
in terms of the Wronskian:
\begin{equation}
  \hat{\mathcal{A}}\psi(x)
  =\frac{\text{W}[\tilde{\phi},\psi](x)}{\tilde{\phi}(x)}
  =\psi^{\text{new}}(x),\quad
  \hat{\mathcal{A}}^{\dagger}\psi^{\text{new}}(x)
  =-\frac{\text{W}[\tilde{\phi}^{-1},\psi^{\text{new}}](x)}
  {\tilde{\phi}^{-1}(x)}
  =(\mathcal{E}-\tilde{\mathcal{E}})\psi(x).
\end{equation}
 
\subsection{Multi-step Darboux transformation}
\label{app:DT_multi}

Assume that the original Hamiltonian $\mathcal{H}=p^2+U(x)$ has eigenstates
$\phi_n(x)$,
\begin{equation}
  \mathcal{H}\phi_n(x)=\mathcal{E}_n\phi_n(x)\quad
  (n=0,1,\ldots),\quad 0=\mathcal{E}_0<\mathcal{E}_1<\cdots,
\end{equation}
where we have chosen the constant term of $U(x)$ such that $\mathcal{E}_0=0$.
We take seed solutions $\tilde{\phi}_{d_j}(x)$,
\begin{equation}
  \mathcal{H}\tilde{\phi}_{d_j}(x)
  =\tilde{\mathcal{E}}_{d_j}\tilde{\phi}_{d_j}(x)
  \quad(j=1,2,\ldots,M).
\end{equation}
By rewriting the original Hamiltonian (0-th step Hamiltonian) as
$\mathcal{H}=\hat{\mathcal{A}}_{d_1}^{\dagger}\hat{\mathcal{A}}_{d_1}
+\tilde{\mathcal{E}}_{d_1}$, we perform the Darboux transformation.
By repeating this procedure, the $s$-step system is obtained from
the $(s-1)$-th step system:
\begin{align}
  &\mathcal{H}_{d_1\ldots d_s}\eqdef\hat{\mathcal{A}}_{d_1\ldots d_s}
  \hat{\mathcal{A}}_{d_1\ldots d_s}^{\dagger}+\tilde{\mathcal{E}}_{d_s}
  \ \bigl(=\hat{\mathcal{A}}_{d_1\ldots d_{s+1}}^{\dagger}
  \hat{\mathcal{A}}_{d_1\ldots d_{s+1}}+\tilde{\mathcal{E}}_{d_{s+1}}
  \text{ for the next step}\bigr),\\
  &\hat{\mathcal{A}}_{d_1\ldots d_s}\eqdef\tfrac{d}{dx}
  -\partial_x\log\bigl|\tilde{\phi}_{d_1\ldots d_s}(x)\bigr|,\quad
  \hat{\mathcal{A}}_{d_1\ldots d_s}^{\dagger}=-\bigl(\tfrac{d}{dx}
  -\partial_x\log\bigl|\tilde{\phi}_{d_1\ldots d_s}^{-1}(x)\bigr|\bigr),\\
  &\phi_{d_1\ldots d_s\,n}(x)\eqdef\hat{\mathcal{A}}_{d_1\ldots d_s}
  \phi_{d_1\ldots d_{s-1}\,n}(x)\quad(n=0,1,\ldots),
  \label{phisn}\\
  &\tilde{\phi}_{d_1\ldots d_s\,\text{v}}(x)\eqdef
  \hat{\mathcal{A}}_{d_1\ldots d_s}
  \tilde{\phi}_{d_1\ldots d_{s-1}\,\text{v}}(x)\quad
  (\text{v}=d_{s+1},d_{s+2},\ldots,d_M),
  \label{phitsv}\\
  &\breve{\Phi}_{d_1\ldots d_s}^{(j)}(x)\eqdef(-1)^{s-j}
  \tilde{\phi}_{d_1\ldots d_{j-1}d_{j+1}\ldots d_s\,d_j}^{\,-1}(x)\quad
  (j=1,2,\ldots,s),
  \label{Phisj}
\end{align}
which satisfy
\begin{align}
  &\mathcal{H}_{d_1\ldots d_s}\phi_{d_1\ldots d_s\,n}(x)
  =\mathcal{E}_n\phi_{d_1\ldots d_s\,n}(x)\quad
  (n=0,1,\ldots),\\
  &\mathcal{H}_{d_1\ldots d_s}\tilde{\phi}_{d_1\ldots d_s\,\text{v}}(x)
  =\tilde{\mathcal{E}}_{\text{v}}\tilde{\phi}_{d_1\ldots d_s\,\text{v}}(x)
  \quad(\text{v}=d_{s+1},d_{s+2},\ldots,d_M),\\
  &\mathcal{H}_{d_1\ldots d_s}\breve{\Phi}_{d_1\ldots d_s}^{(j)}(x)
  =\tilde{\mathcal{E}}_{d_j}\breve{\Phi}_{d_1\ldots d_s}^{(j)}(x)\quad
  (j=1,2,\ldots,s),\\
  &\hat{\mathcal{A}}_{d_1\ldots d_s}^{\dagger}\phi_{d_1\ldots d_s\,n}(x)
  =(\mathcal{E}_n-\tilde{\mathcal{E}}_{d_s})\phi_{d_1\ldots d_{s-1}\,n}(x).
  \label{Ahsphisn=}
\end{align}
The wavefunctions \eqref{phisn}--\eqref{Phisj} and the potential
($\mathcal{H}_{d_1\ldots d_s}=p^2+U_{d_1\ldots d_s}(x)$) are expressed
in terms of the Wronskian,
\begin{align}
  &\phi_{d_1\ldots d_s\,n}(x)
  =\frac{\text{W}[\tilde{\phi}_{d_1},\ldots,\tilde{\phi}_{d_s},\phi_n](x)}
  {\text{W}[\tilde{\phi}_{d_1},\ldots,\tilde{\phi}_{d_s}](x)},
  \label{phisn=W/W}\\
  &\tilde{\phi}_{d_1\ldots d_s\,\text{v}}(x)
  =\frac{\text{W}[\tilde{\phi}_{d_1},\ldots,\tilde{\phi}_{d_s},
  \tilde{\phi}_{\text{v}}](x)}
  {\text{W}[\tilde{\phi}_{d_1},\ldots,\tilde{\phi}_{d_s}](x)},
  \label{phitsv=W/W}\\
  &\breve{\Phi}_{d_1\ldots d_s}^{(j)}(x)
  =\frac{\text{W}[\tilde{\phi}_{d_1},\ldots,\tilde{\phi}_{d_{j-1}},
  \tilde{\phi}_{d_{j+1}},\ldots,\tilde{\phi}_{d_s}](x)}
  {\text{W}[\tilde{\phi}_{d_1},\ldots,\tilde{\phi}_{d_s}](x)},
  \label{Phisj=W/W}\\
  &U_{d_1\ldots d_s}(x)=U(x)-2\partial_x^2\log
  \bigl|\text{W}[\tilde{\phi}_{d_1},\ldots,\tilde{\phi}_{d_s}](x)\bigr|,
\end{align}
which are shown by using \eqref{Wformula1}--\eqref{Wformula2}.
Note that $\mathcal{H}_{d_1\ldots d_s}$, $\hat{\mathcal{A}}_{d_1\ldots d_s}$
and $\hat{\mathcal{A}}_{d_1\ldots d_s}^{\dagger}$ are independent of the
order of $d_1,\ldots,d_s$ ($\phi_{d_1\ldots d_s\,n}(x)$,
$\tilde{\phi}_{d_1\ldots d_s\,\text{v}}(x)$ and
$\breve{\Phi}_{d_1\ldots d_s}^{(j)}(x)$ may change sign).
Let us define $\hat{\mathcal{A}}^{(d_1\ldots d_s)}$ as
\begin{equation}
  \hat{\mathcal{A}}^{(d_1\ldots d_s)}\eqdef
  \hat{\mathcal{A}}_{d_1\dots d_s}\cdots\hat{\mathcal{A}}_{d_1d_2}
  \hat{\mathcal{A}}_{d_1},\quad
  \hat{\mathcal{A}}^{(d_1\ldots d_s)\,\dagger}=
  \hat{\mathcal{A}}_{d_1}^{\dagger}\hat{\mathcal{A}}_{d_1d_2}^{\dagger}
  \cdots\hat{\mathcal{A}}_{d_1\dots d_s}^{\dagger}.
\end{equation}
Then we have
\begin{equation}
  \hat{\mathcal{A}}^{(d_1\ldots d_s)}\phi_n(x)=\phi_{d_1\ldots d_s\,n}(x),
  \quad
  \hat{\mathcal{A}}^{(d_1\ldots d_s)\,\dagger}\phi_{d_1\ldots d_s\,n}(x)
  =\prod_{j=1}^s(\mathcal{E}_n-\tilde{\mathcal{E}}_{d_j})\cdot
  \phi_n(x).
  \label{Ahsphin}
\end{equation}
They are expressed in terms of the Wronskian,
\begin{align}
  &\hat{\mathcal{A}}^{(d_1\ldots d_s)}\phi_n(x)
  =\frac{\text{W}[\tilde{\phi}_{d_1},\ldots,\tilde{\phi}_{d_s},\phi_n](x)}
  {\text{W}[\tilde{\phi}_{d_1},\ldots,\tilde{\phi}_{d_s}](x)}
  =\phi_{d_1\ldots d_s\,n}(x),
  \label{Ahsphin=W/W}\\
  &\hat{\mathcal{A}}^{(d_1\ldots d_s)\,\dagger}\phi_{d_1\ldots d_s\,n}(x)
  =(-1)^s\frac{\text{W}[\breve{\Phi}_{d_1\ldots d_s}^{(1)},\ldots,
  \breve{\Phi}_{d_1\ldots d_s}^{(s)},\phi_{d_1\ldots d_s\,n}](x)}
  {\text{W}[\breve{\Phi}_{d_1\ldots d_s}^{(1)},\ldots,
  \breve{\Phi}_{d_1\ldots d_s}^{(s)}](x)}
  =\prod_{j=1}^s(\mathcal{E}_n-\tilde{\mathcal{E}}_{d_j})\cdot
  \phi_n(x)\n
  &\phantom{
  \hat{\mathcal{A}}^{(d_1\ldots d_s)\,\dagger}\phi_{d_1\ldots d_s\,n}(x)}
  =(-1)^{\frac12s(s+1)}
  \frac{\text{W}\bigl[w_1,\ldots,w_s,
  \text{W}[\tilde{\phi}_{d_1},\ldots,\tilde{\phi}_{d_s},\phi_n]\bigr](x)}
  {\text{W}[\tilde{\phi}_{d_1},\ldots,\tilde{\phi}_{d_s}](x)^{s}},
  \label{Ahdsphisn=W/W}
\end{align}
where $w_j$ is
$w_j(x)=\text{W}[\tilde{\phi}_{d_1},\ldots,\tilde{\phi}_{d_{j-1}},
\tilde{\phi}_{d_{j+1}},\ldots,\tilde{\phi}_{d_s}](x)$.
As far as we know, this formula \eqref{Ahdsphisn=W/W} is new.
Eq.\,\eqref{Ahsphin=W/W} is already given in \eqref{phisn=W/W}.
{}To obtain the second line of \eqref{Ahdsphisn=W/W} from the first line,
we use \eqref{Wformula4}, and the formula \eqref{Ahdsphisn=W/W} is shown
by using \eqref{Wformula1}--\eqref{Wformula2}.

\subsection{Polynomial type solutions}
\label{app:DT_poly}

Let us assume that eigenfunctions $\phi_n(x)$ and seed solutions
$\tilde{\phi}_{\text{v}}(x)$ are polynomial type solutions, namely
they have the following forms,
\begin{equation}
  \phi_n(x)=\phi_0(x)P_n\bigl(\eta(x)\bigr),\quad
  \tilde{\phi}_{\text{v}}(x)=\tilde{\phi}_{0(\text{v})}(x)
  \xi_{\text{v}}\bigl(\eta(x)\bigr),
\end{equation}
where $\phi_0(x)$, $\tilde{\phi}_{0(\text{v})}(x)$ and $\eta(x)$ are
functions of $x$, and $P_n(\eta)$ and $\xi_{\text{v}}(\eta)$ are polynomials
in $\eta$.
For concrete examples, e.g. Laguerre and Jacobi cases given in
Appendix\,\ref{app:JL}, the Wronskians in \eqref{Ahsphin=W/W} have the
following forms,
\begin{align}
  \text{W}[\tilde{\phi}_{d_1},\ldots,\tilde{\phi}_{d_s}](x)
  &=\text{(some function of $x$)}\times
  \Xi_{d_1\ldots d_s}\bigl(\eta(x)\bigr),
  \label{W=Xi}\\
  \text{W}[\tilde{\phi}_{d_1},\ldots,\tilde{\phi}_{d_s},\phi_n](x)
  &=\text{(some function of $x$)}\times
  P_{d_1\ldots d_s,n}\bigl(\eta(x)\bigr),
  \label{W=Psn}
\end{align}
where $\Xi_{d_1\ldots d_s}(\eta)$ and $P_{d_1\ldots d_s,n}(\eta)$ are
polynomials in $\eta$.
Therefore $\phi_{d_1\ldots d_s\,n}(x)$ has the following form,
\begin{equation}
  \phi_{d_1\ldots d_s\,n}(x)
  =\Psi_{d_1\ldots d_s}(x)P_{d_1\dots d_s,n}\bigl(\eta(x)\bigr),\quad
  \Psi_{d_1\ldots d_s}(x)=\frac{\text{(some function of $x$)}}
  {\Xi_{d_1\ldots d_s}\bigl(\eta(x)\bigr)}.
\end{equation}
Let us define the step forward ($\hat{\mathcal{F}}$) and backward
($\hat{\mathcal{B}}$) shift operators as,
\begin{align}
  \hat{\mathcal{F}}_{d_1\ldots d_s}&\eqdef
  \Psi_{d_1\ldots d_s}^{\,-1}(x)\circ\hat{\mathcal{A}}_{d_1\ldots d_s}\circ
  \Psi_{d_1\ldots d_{s-1}}(x),
  \label{Fhs}\\
  \hat{\mathcal{B}}_{d_1\ldots d_s}&\eqdef
  \Psi_{d_1\ldots d_{s-1}}^{\,-1}(x)\circ
  \hat{\mathcal{A}}_{d_1\ldots d_s}^{\dagger}\circ
  \Psi_{d_1\ldots d_s}(x),
  \label{Bhs}
\end{align}
where $\Psi_{d_1\ldots d_{s-1}}(x)\bigl|_{s=1}=\phi_0(x)$.
The relations \eqref{phisn} and \eqref{Ahsphisn=} are rewritten as
\begin{equation}
  \hat{\mathcal{F}}_{d_1\ldots d_s}P_{d_1\ldots d_{s-1},n}(\eta)
  =P_{d_1\ldots d_s,n}(\eta),\quad
  \hat{\mathcal{B}}_{d_1\ldots d_s}P_{d_1\ldots d_s,n}(\eta)
  =(\mathcal{E}_n-\tilde{\mathcal{E}}_{d_s})P_{d_1\ldots d_{s-1},n}(\eta).
  \label{FhsPsn=}
\end{equation}
The relations \eqref{Ahsphin} are also rewritten as
\begin{equation}
  \hat{\mathcal{F}}^{(d_1\ldots d_s)}P_n(\eta)
  =P_{d_1\ldots d_s,n}(\eta),\quad
  \hat{\mathcal{B}}^{(d_1\ldots d_s)}P_{d_1\ldots d_s,n}(\eta)
  =\prod_{j=1}^s(\mathcal{E}_n-\tilde{\mathcal{E}}_{d_j})\cdot
  P_n(\eta),
  \label{FhsPn=}
\end{equation}
where the multi-step forward and backward shift operators,
$\hat{\mathcal{F}}^{(d_1\ldots d_s)}$ and
$\hat{\mathcal{B}}^{(d_1\ldots d_s)}$, are defined by
\begin{align}
  &\hat{\mathcal{F}}^{(d_1\ldots d_s)}\eqdef
  \hat{\mathcal{F}}_{d_1\dots d_s}\cdots\hat{\mathcal{F}}_{d_1d_2}
  \hat{\mathcal{F}}_{d_1}
  =\Psi_{d_1\ldots d_s}^{\,-1}(x)\circ\hat{\mathcal{A}}^{(d_1\ldots d_s)}
  \circ\phi_0(x),
  \label{Fh(s)}\\
  &\hat{\mathcal{B}}^{(d_1\ldots d_s)}\eqdef
  \hat{\mathcal{B}}_{d_1}\hat{\mathcal{B}}_{d_1d_2}
  \cdots\hat{\mathcal{B}}_{d_1\dots d_s}
  =\phi_0^{\,-1}(x)\circ\hat{\mathcal{A}}^{(d_1\ldots d_s)\,\dagger}\circ
  \Psi_{d_1\ldots d_s}(x).
  \label{Bh(s)}
\end{align}
We have
\begin{align}
  \hat{\mathcal{B}}^{(d_1\ldots d_s)}\hat{\mathcal{F}}^{(d_1\ldots d_s)}
  P_n(\eta)
  &=\prod_{j=1}^s(\mathcal{E}_n-\tilde{\mathcal{E}}_{d_j})\cdot P_n(\eta),\\
  \hat{\mathcal{F}}^{(d_1\ldots d_s)}\hat{\mathcal{B}}^{(d_1\ldots d_s)}
  P_{d_1\ldots d_s,n}(\eta)
  &=\prod_{j=1}^s(\mathcal{E}_n-\tilde{\mathcal{E}}_{d_j})\cdot
  P_{d_1\ldots d_s,n}(\eta).
  \label{BhsFhsPn=}
\end{align}
Remark that
\begin{align}
  \hat{\mathcal{F}}^{(d_1\ldots d_s)}P_n\bigl(\eta(x)\bigr)
  &=\Psi_{d_1\ldots d_s}^{\,-1}(x)\hat{\mathcal{A}}^{(d_1\ldots d_s)}
  \phi_n(x),
  \label{FhsPn}\\
  \hat{\mathcal{B}}^{(d_1\ldots d_s)}P_{d_1\ldots d_s,n}\bigl(\eta(x)\bigr)
  &=\phi_0^{\,-1}(x)\hat{\mathcal{A}}^{(d_1\ldots d_s)\,\dagger}
  \phi_{d_1\ldots d_s\,n}(x).
  \label{BhsPsn}
\end{align}

\section{Multi-indexed Laguerre and Jacobi polynomials}
\label{app:JL}

In this appendix we review the multi-indexed orthogonal polynomials of
Laguerre and Jacobi types \cite{os25}, mainly their algebraic properties.
They are obtained from the Laguerre and Jacobi polynomials by applying
the multi-step Darboux transformation explained in Appendix\,\ref{app:DT}.
If necessary, we write the parameter $\bm{\lambda}$ dependence explicitly,
$\hat{\mathcal{A}}_{d_1\ldots d_s}
=\hat{\mathcal{A}}_{d_1\ldots d_s}(\bm{\lambda})$,
$\phi_{d_1\ldots d_s\,n}(x)=\phi_{d_1\ldots d_s\,n}(x;\bm{\lambda})$,
$P_{d_1\ldots d_s,n}(\eta)=P_{d_1\ldots d_s,n}(\eta;\bm{\lambda})$,
$\mathcal{E}_n=\mathcal{E}_n(\bm{\lambda})$, etc.
We assume that the parameters ($g$ and $h$) are generic such that
$c_{d_1\ldots d_s}^{\Xi}\neq 0$ \eqref{cXis},
$c_{d_1\ldots d_s,n}^{P}\neq 0$ \eqref{cPsn} and
$\mathcal{E}_n-\tilde{\mathcal{E}}_{d_j}\neq 0$.

The original systems are the radial oscillator and the Darboux-P\"oschl-Teller
potential for Laguerre (L) and Jacobi (J) cases respectively:
\begin{align}
  \text{L}:\ &
  \mathcal{H}=p^2+x^2+\frac{g(g-1)}{x^2}-2g-1,\quad
  0<x<\infty,\quad g>\tfrac12,\n
  &\bm{\lambda}=g,\quad\bm{\delta}=1,
  \quad\cF=2,\quad\mathcal{E}_n=4n,\n
  &\eta(x)=x^2,\quad\phi_0(x)=e^{-\frac12x^2}x^g,\quad
  P_n(\eta)=L^{(g-\frac12)}_n(\eta),\\
  \text{J}:\ &
  \mathcal{H}=p^2+\frac{g(g-1)}{\sin x^2}+\frac{h(h-1)}{\cos x^2}
  -(g+h)^2,\quad
  0<x<\tfrac{\pi}{2},\quad g,h>\tfrac12,\n
  &\bm{\lambda}=(g,h),\quad
  \bm{\delta}=(1,1),\quad\cF=-4,\quad
  \mathcal{E}_n=4n(n+g+h),\n
  &\eta(x)=\cos2x,\quad
  \phi_0(x)=(\sin x)^g(\cos x)^h,\quad
  P_n(\eta)=P^{(g-\frac12,h-\frac12)}_n(\eta),
\end{align}
with $\phi_n(x)=\phi_0(x)P_n\bigl(\eta(x)\bigr)$. 
We take the virtual state wavefunctions as seed solutions,
\begin{align}
  \text{L}:\ &\text{type $\I$: }
  \tilde{\phi}^{\I}_{\text{v}}(x;\bm{\lambda})
  \eqdef i^{-g}\phi_{\text{v}}(ix;\bm{\lambda})
  =e^{\frac12x^2}x^gL^{(g-\frac12)}_{\text{v}}\bigl(-\eta(x)\bigr),\quad
  \tilde{\bm{\delta}}^{\I}\eqdef 1,\n
  &\phantom{\text{type $\I$: }}
  \tilde{\mathcal{E}}^{\I}_{\text{v}}=-4(g+\text{v}+\tfrac12),\\
  &\text{type $\II$: }
  \tilde{\phi}^{\II}_{\text{v}}(x;\bm{\lambda})
  \eqdef \phi_{\text{v}}\bigl(x;\mathfrak{t}^{\II}(\bm{\lambda})\bigr)
  =e^{-\frac12x^2}x^{1-g}L^{(\frac12-g)}_{\text{v}}\bigl(\eta(x)\bigr),\quad
  \tilde{\bm{\delta}}^{\II}\eqdef-1,\n
  &\phantom{\text{type $\II$: }}
  \mathfrak{t}^{\II}(\bm{\lambda})\eqdef 1-g,\quad
  \tilde{\mathcal{E}}^{\II}_{\text{v}}(\bm{\lambda})
  =-4(g-\text{v}-\tfrac12),\\
  \text{J}:\ &\text{type $\I$: }
  \tilde{\phi}^{\I}_{\text{v}}(x;\bm{\lambda})
  \eqdef\phi_{\text{v}}\bigl(x;\mathfrak{t}^{\I}(\bm{\lambda})\bigr)
  =(\sin x)^g(\cos x)^{1-h}P^{(g-\frac12,\frac12-h)}_{\text{v}}
  \bigl(\eta(x)\bigr),
  \quad\tilde{\bm{\delta}}^{\I}\eqdef(1,-1),\n
  &\phantom{\text{type $\I$: }}
  \mathfrak{t}^{\I}(\bm{\lambda})=(g,1-h),\quad
  \tilde{\mathcal{E}}^{\I}_{\text{v}}
  =-4(g+\text{v}+\tfrac12)(h-\text{v}-\tfrac12),\\
  &\text{type $\II$: }
  \tilde{\phi}^{\II}_{\text{v}}(x;\bm{\lambda})
  \eqdef\phi_{\text{v}}\bigl(x;\mathfrak{t}^{\II}(\bm{\lambda})\bigr)
  =(\sin x)^{1-g}(\cos x)^hP^{(\frac12-g,h-\frac12)}_{\text{v}}
  \bigl(\eta(x)\bigr),
  \quad\tilde{\bm{\delta}}^{\II}\eqdef(-1,1),\n
  &\phantom{\text{type $\II$: }}
  \mathfrak{t}^{\II}(\bm{\lambda})=(1-g,h),\quad
  \tilde{\mathcal{E}}^{\II}_{\text{v}}
  =-4(g-\text{v}-\tfrac12)(h+\text{v}+\tfrac12),
\end{align}
where the range of $\text{v}$, $g$, $h$ are found in \cite{os25}.
These virtual state wavefunctions are labeled by the degree $\text{v}$ of
the polynomial part and the type $\text{t}$ ($\I$ or $\II$),
$(\text{v},\text{t})$ which we write as $\text{v}^{\text{t}}$.
For simplicity in notation, we abbreviate $\text{v}^{\text{t}}$ as
$\text{v}$ in most places.

Eqs.\,\eqref{W=Xi}--\eqref{W=Psn} become
\begin{align}
  &\text{W}[\tilde{\phi}_{d_1},\ldots,\tilde{\phi}_{d_s}](x)
  =\cF^{\frac12s(s-1)}\Xi_{d_1\ldots d_s}(\eta)
  \times\left\{
  \begin{array}{ll}
  \eta^{s'(s'+g-\frac12)}e^{s'\eta}&:\text{L}\\
  \bigl(\frac{1-\eta}{2}\bigr)^{s'(s'+g-\frac12)}
  \bigl(\frac{1+\eta}{2}\bigr)^{-s'(-s'+h-\frac12)}
  &:\text{J}
  \end{array}\right.,
  \label{W=Xi2}\\
  &\text{W}[\tilde{\phi}_{d_1},\ldots,\tilde{\phi}_{d_s},\phi_n](x)\n
  &\quad=\cF^{\frac12s(s+1)}P_{d_1\ldots d_s,n}(\eta)
  \times\left\{
  \begin{array}{ll}
  \eta^{(s'+\frac12)(s'+g)}e^{(s'-\frac12)\eta}&:\text{L}\\
  \bigl(\frac{1-\eta}{2}\bigr)^{(s'+\frac12)(s'+g)}
  \bigl(\frac{1+\eta}{2}\bigr)^{(-s'+\frac12)(-s'+h)}
  &:\text{J}
  \end{array}\right.,
  \label{W=Psn2}
\end{align}
where $\eta=\eta(x)$, $s'=\frac12(s_{\I}-s_{\II})$ and
$s_{\text{t}}=\#\{d_j\,|\,d_j\text{\,:\ type $\text{t}$},\,j=1,\ldots,s\}$
($\text{t}=\I,\II$).
Here the denominator polynomial $\Xi_{d_1\ldots d_s}(\eta)$ and
the multi-indexed orthogonal polynomial
$P_{d_1\ldots d_s,n}(\eta)$, which are polynomials of degree
$\ell_{d_1\ldots d_s}$ and $\ell_{d_1\ldots d_s}+n$ in $\eta$ respectively,
are defined by
\begin{align}
  \Xi_{d_1\ldots d_s}(\eta)
  &\eqdef\text{W}[\mu_{d_1},\ldots,\mu_{d_s}](\eta)\times\left\{
  \begin{array}{ll}
  \eta^{(s_{\I}+g-\frac12)s_{\II}}e^{-s_{\I}\eta}&:\text{L}\\
  \bigl(\frac{1-\eta}{2}\bigr)^{(s_{\I}+g-\frac12)s_{\II}}
  \bigl(\frac{1+\eta}{2}\bigr)^{(s_{\II}+h-\frac12)s_{\I}}&:\text{J}
  \end{array}\right.,
  \label{Xis}\\
  P_{d_1\ldots d_s,n}(\eta)
  &\eqdef\text{W}[\mu_{d_1},\ldots,\mu_{d_s},P_n](\eta)\times\left\{
  \begin{array}{ll}
  \eta^{(s_{\I}+g+\frac12)s_{\II}}e^{-s_{\I}\eta}&:\text{L}\\
  \bigl(\frac{1-\eta}{2}\bigr)^{(s_{\I}+g+\frac12)s_{\II}}
  \bigl(\frac{1+\eta}{2}\bigr)^{(s_{\II}+h+\frac12)s_{\I}}&:\text{J}
  \end{array}\right.,
  \label{Psn}\\
  \mu_{\text{v}}(\eta)&=\left\{
  \begin{array}{ll}
  e^{\eta}\times L^{(g-\frac12)}_{\text{v}}(-\eta)
  &:\text{L, $\text{v}$ type $\I$}\\
  \eta^{\frac12-g}\times L^{(\frac12-g)}_{\text{v}}(\eta)
  &:\text{L, $\text{v}$ type $\II$}\\
  \bigl(\frac{1+\eta}{2}\bigr)^{\frac12-h}\times
  P^{(g-\frac12,\frac12-h)}_{\text{v}}(\eta)
  &:\text{J, $\text{v}$ type $\I$}\\[3pt]
  \bigl(\frac{1-\eta}{2}\bigr)^{\frac12-g}\times
  P^{(\frac12-g,h-\frac12)}_{\text{v}}(\eta)
  &:\text{J, $\text{v}$ type $\II$}
  \end{array}\right.,
  \label{muv}
\end{align}
and
\begin{equation}
  \ell_{d_1\ldots d_s}\eqdef\sum_{j=1}^sd_j-\frac12s(s-1)+2s_{\I}s_{\II}.
  \label{ells}
\end{equation}
Since $L^{(\alpha)}_n(\eta)$ and $P^{(\alpha,\beta)}_n(\eta)$ belong to
$\mathbb{Q}[\eta,\alpha,\beta]$, these polynomials $\Xi_{d_1\ldots d_s}(\eta)$
and $P_{d_1\ldots d_s,n}(\eta)$ also belong to $\mathbb{Q}[\eta,g,h]$.
Under a permutation of $d_j$'s, $\Xi_{d_1\ldots d_s}(\eta)$ and
$P_{d_1\ldots d_s,n}(\eta)$ change their overall sign,
$\Xi_{d_{\sigma_1}\ldots d_{\sigma_s}}(\eta)
=\text{sgn}\genfrac{(}{)}{0pt}{}{1\ \ldots\ s\ }{\sigma_1\,\ldots\,\sigma_s}
\,\Xi_{d_1\ldots d_s}(\eta)$ and
$P_{d_{\sigma_1}\ldots d_{\sigma_s},n}(\eta)
=\text{sgn}\genfrac{(}{)}{0pt}{}{1\ \ldots\ s\ }{\sigma_1\,\ldots\,\sigma_s}
P_{d_1\ldots d_s,n}(\eta)$.
We denote the coefficients of the highest degree term of the polynomials
$\Xi_{d_1\ldots d_s}(\eta)$ and $P_{d_1\ldots d_s,n}(\eta)$ as
\begin{align}
  &\Xi_{d_1\ldots d_s}(\eta)
  =c_{d_1\ldots d_s}^{\Xi}\eta^{\ell_{d_1\ldots d_s}}
  +(\text{lower order terms}),\n
  &P_{d_1\ldots d_s,n}(\eta)
  =c_{d_1\ldots d_s,n}^{P}\eta^{\ell_{d_1\ldots d_s}+n}
  +(\text{lower order terms}).
\end{align}
In the `standard order'
$\{d^{\I}_1,\ldots,d^{\I}_{s_{\I}},d^{\II}_1,\ldots,d^{\II}_{s_{\II}}\}$,
these coefficients are \cite{equiv_miop}
\begin{align}
  c_{d^{\I}_1\ldots d^{\II}_{s_{\II}}}^{\Xi}&=
  \prod_{j=1}^{s_{\I}}c^{\I}_{d^{\I}_j}\cdot
  \prod_{j=1}^{s_{\II}}c^{\II}_{d^{\II}_j}\cdot
  \prod_{1\leq j<k\leq s_{\I}}(d^{\I}_k-d^{\I}_j)\cdot
  \prod_{1\leq j<k\leq s_{\II}}(d^{\II}_k-d^{\II}_j)\n
  &\quad\times\left\{
  \begin{array}{ll}
  (-1)^{s_{\I}s_{\II}}
  &:\text{L}\\
  \prod\limits_{j=1}^{s_{\I}}\prod\limits_{k=1}^{s_{\II}}
  \frac14(g-h+d^{\I}_j-d^{\II}_k)
  &:\text{J}
  \end{array}\right.,
  \label{cXis}\\
  c_{d^{\I}_1\ldots d^{\II}_{s_{\II}},n}^{P}&=
  c_{d^{\I}_1\ldots d^{\II}_{s_{\II}}}^{\Xi}c_n
  \times\left\{
  \begin{array}{ll}
  (-1)^{s_{\I}}\prod\limits_{j=1}^{s_{\II}}(g+n-d^{\II}_j-\frac12)
  &:\text{L}\\
  \prod\limits_{j=1}^{s_{\I}}\frac12(h+n-d^{\I}_j-\frac12)\cdot
  \prod\limits_{j=1}^{s_{\II}}\frac{-1}{2}(g+n-d^{\II}_j-\frac12)
  &:\text{J}
  \end{array}\right.,
  \label{cPsn}
\end{align}
where $c_n$, $c^{\I}_{\text{v}}$ and $c^{\II}_{\text{v}}$ are
\begin{align}
  &P_n(\eta)=c_n\eta^n+(\text{lower order terms}),\quad
  c_n=\left\{
  \begin{array}{ll}
  {\displaystyle\frac{(-1)^n}{n!}}&:\text{L}\\[5pt]
  {\displaystyle\frac{(n+g+h)_n}{2^n\,n!}}&:\text{J}
  \end{array}\right.,\\
  &c^{\I}_{\text{v}}(\bm{\lambda})\eqdef\left\{
  \begin{array}{ll}
  (-1)^{\text{v}}c_{\text{v}}(\bm{\lambda})&:\text{L}\\[2pt]
  c_{\text{v}}\bigl(\mathfrak{t}^{\I}(\bm{\lambda})\bigr)&:\text{J}
  \end{array}\right.,\quad
  c^{\II}_{\text{v}}(\bm{\lambda})\eqdef
  c_{\text{v}}\bigl(\mathfrak{t}^{\II}(\bm{\lambda})\bigr)
  \ :\text{L,\,J}.
\end{align}

{}From \eqref{phisn=W/W} and \eqref{W=Xi2}--\eqref{W=Psn2}, we obtain
\begin{align}
  &\phi_{d_1\ldots d_s\,n}(x;\bm{\lambda})
  =\Psi_{d_1\ldots d_s}(x;\bm{\lambda})
  P_{d_1\ldots d_s,n}\bigl(\eta(x);\bm{\lambda}\bigr),\n
  &\Psi_{d_1\ldots d_s}(x;\bm{\lambda})
  =\cF^s\frac{\phi_0(x;\bm{\lambda}^{[s_{\I},s_{\II}]})}
  {\Xi_{d_1\ldots d_s}\bigl(\eta(x);\bm{\lambda}\bigr)},\quad
  \bm{\lambda}^{[s_{\I},s_{\II}]}\eqdef
  \bm{\lambda}+s_{\I}\tilde{\bm{\delta}}^{\I}
  +s_{\II}\tilde{\bm{\delta}}^{\II}.
\end{align}
Explicit forms of the step forward and backward shift operators
$\hat{\mathcal{F}}_{d_1\ldots d_s}$ and
$\hat{\mathcal{B}}_{d_1\ldots d_s}$ are given by eqs.(A.1)--(A.4)
in \cite{rrmiop2}.
To calculate the multi-step one $\hat{\mathcal{F}}^{(d_1\ldots d_s)}$ and
$\hat{\mathcal{B}}^{(d_1\ldots d_s)}$, however, they are not so useful.
Instead, we use \eqref{Ahsphin=W/W}--\eqref{Ahdsphisn=W/W} and
\eqref{FhsPn}--\eqref{BhsPsn}.
By using \eqref{W=Xi2}--\eqref{W=Psn2}, we obtain
\begin{align}
  \hat{\mathcal{F}}^{(d_1\ldots d_s)}P_n(\eta)
  &=\rho^{(d_1\ldots d_s)}_{\hat{\mathcal{F}}}(\eta)
  \text{W}[\mu_{d_1},\ldots,\mu_{d_s},P_n](\eta),
  \label{Fh(s)Pn}\\
  \hat{\mathcal{B}}^{(d_1\ldots d_s)}P_{d_1\ldots d_s,n}(\eta)
  &=\rho^{(d_1\ldots d_s)}_{\hat{\mathcal{B}}}(\eta)
  \text{W}[m_1,\ldots,m_s,P_{d_1\ldots d_s,n}](\eta).
  \label{Bh(s)Psn}
\end{align}
Here $\mu_{\text{v}}(\eta)$ is given in \eqref{muv}, and
$\rho^{(d_1\ldots d_s)}_{\hat{\mathcal{F}}}(\eta)$,
$\rho^{(d_1\ldots d_s)}_{\hat{\mathcal{B}}}(\eta)$
and $m_j(\eta)$ are
\begin{align}
  &\rho^{(d_1\ldots d_s)}_{\hat{\mathcal{F}}}(\eta)
  \eqdef\left\{
  \begin{array}{ll}
  \eta^{(s_{\I}+g+\frac12)s_{\II}}e^{-s_{\I}\eta}&:\text{L}\\
  \bigl(\frac{1-\eta}{2}\bigr)^{(s_{\I}+g+\frac12)s_{\II}}
  \bigl(\frac{1+\eta}{2}\bigr)^{(s_{\II}+h+\frac12)s_{\I}}&:\text{J}
  \end{array}\right.,
  \label{rhoFh}\\
  &\rho^{(d_1\ldots d_s)}_{\hat{\mathcal{B}}}(\eta)
  \eqdef\frac{\cF^{2s}(-1)^{\frac12s(s+1)}}{\Xi_{d_1\dots d_s}(\eta)^s}
  \times\left\{
  \begin{array}{ll}
  \eta^{s_{\I}(s_{\I}+g+\frac12)}e^{-s_{\II}\eta}&:\text{L}\\
  \bigl(\frac{1-\eta}{2}\bigr)^{s_{\I}(s_{\I}+g+\frac12)}
  \bigl(\frac{1+\eta}{2}\bigr)^{s_{\II}(s_{\II}+h+\frac12)}&:\text{J}
  \end{array}\right.,
  \label{rhoBh}\\
  &m_j(\eta)=m^{(d_1\ldots d_s)}_j(\eta)\eqdef
  \Xi_{d_1\ldots d_{j-1}d_{j+1}\dots d_s}(\eta)\times\left\{
  \begin{array}{ll}
  \eta^{-(s_{\I}-s_{\II}+g-\frac12)}&:\text{L, $d_j$ type $\I$}\\
  e^{\eta}&:\text{L, $d_j$ type $\II$}\\
  \bigl(\frac{1-\eta}{2}\bigr)^{-(s_{\I}-s_{\II}+g-\frac12)}
  &:\text{J, $d_j$ type $\I$}\\
  \bigl(\frac{1+\eta}{2}\bigr)^{-(s_{\II}-s_{\I}+h-\frac12)}
  &:\text{J, $d_j$ type $\II$}
  \end{array}\right.,
  \label{mj}
\end{align}
where $\Xi_{d_1\ldots d_{j-1}d_{j+1}\dots d_s}(\eta)\bigl|_{s=1}=1$.
This formula \eqref{Bh(s)Psn} (see \eqref{FhsPn=}) is new.
For the exceptional Hermite polynomial with multi indices, the formula like 
\eqref{Bh(s)Psn} was given in \cite{gkkm15}.

The Hamiltonian $\mathcal{H}_{d_1\ldots d_s}$ can be written in the
standard form:
\begin{equation}
  \mathcal{H}_{d_1\ldots d_s}
  =\mathcal{A}_{d_1\ldots d_s}^{\dagger}\mathcal{A}_{d_1\ldots d_s},\quad
  \mathcal{A}_{d_1\ldots d_s}\eqdef\tfrac{d}{dx}
  -\partial_x\log\bigl|\phi_{d_1\ldots d_s\,0}(x)\bigr|.
\end{equation}
The shape invariance of the original system is inherited by the deformed
system,
\begin{equation}
  \mathcal{A}_{d_1\ldots d_s}(\bm{\lambda})
  \mathcal{A}_{d_1\ldots d_s}(\bm{\lambda})^{\dagger}
  =\mathcal{A}_{d_1\ldots d_s}(\bm{\lambda}+\bm{\delta})^{\dagger}
  \mathcal{A}_{d_1\ldots d_s}(\bm{\lambda}+\bm{\delta})
  +\mathcal{E}_1(\bm{\lambda}).
  \label{shapeinv}
\end{equation}
As a consequence of the shape invariance, we have
\begin{align}
  \mathcal{A}_{d_1\ldots d_s}(\bm{\lambda})
  \phi_{d_1\ldots d_s\,n}(x;\bm{\lambda})
  &=f_n(\bm{\lambda})
  \phi_{d_1\ldots d_s\,n-1}(x;\bm{\lambda}+\bm{\delta}),
  \label{Asphisn=}\\
  \mathcal{A}_{d_1\ldots d_s}(\bm{\lambda})^{\dagger}
  \phi_{d_1\ldots d_s\,n-1}(x;\bm{\lambda}+\bm{\delta})
  &=b_{n-1}(\bm{\lambda})
  \phi_{d_1\ldots d_s\,n}(x;\bm{\lambda}),
  \label{Adsphisn=}
\end{align}
where the constants $f_n(\bm{\lambda})$ and $b_{n-1}(\bm{\lambda})$ are
the factors of the eigenvalue
$f_n(\bm{\lambda})b_{n-1}(\bm{\lambda})=\mathcal{E}_n(\bm{\lambda})$:
\begin{align}
  &f_n(\bm{\lambda})=\left\{
  \begin{array}{ll}
  -2&:\text{L}\\
  -2(n+g+h)&:\text{J}
  \end{array}\right.,
  \quad
  b_{n-1}(\bm{\lambda})=-2n\ \ :\text{L,\,J}.
\end{align}
The relations \eqref{Asphisn=}--\eqref{Adsphisn=} give the forward and
backward shift relations,
\begin{align}
  \mathcal{F}_{d_1\ldots d_s}(\bm{\lambda})
  P_{d_1\ldots d_s,n}(\eta;\bm{\lambda})
  &=f_n(\bm{\lambda})
  P_{d_1\ldots d_s,n-1}(\eta;\bm{\lambda}+\bm{\delta}),
  \label{FsPsn=}\\
  \mathcal{B}_{d_1\ldots d_s}(\bm{\lambda})
  P_{d_1\ldots d_s,n-1}(\eta;\bm{\lambda}+\bm{\delta})
  &=b_{n-1}(\bm{\lambda})
  P_{d_1\ldots d_s,n}(\eta;\bm{\lambda}),
  \label{BsPsn=}
\end{align}
where the forward ($\mathcal{F}$) and backward ($\mathcal{B}$) shift
operators are defined by
\begin{align}
  \mathcal{F}_{d_1\ldots d_s}(\bm{\lambda})&\eqdef
  \Psi_{d_1\ldots d_s}^{\,-1}(x;\bm{\lambda}+\bm{\delta})\circ
  \mathcal{A}_{d_1\ldots d_s}(\bm{\lambda})\circ
  \Psi_{d_1\ldots d_s}(x;\bm{\lambda}),
  \label{Fsdef}\\
  \mathcal{B}_{d_1\ldots d_s}(\bm{\lambda})&\eqdef
  \Psi_{d_1\ldots d_s}^{\,-1}(x;\bm{\lambda})\circ
  \mathcal{A}_{d_1\ldots d_s}(\bm{\lambda})^{\dagger}\circ
  \Psi_{d_1\ldots d_s}(x;\bm{\lambda}+\bm{\delta}).
  \label{Bsdef}
\end{align}
 Another consequence of the shape invariance is the following proportionality,
\begin{align}
  P_{d_1\ldots d_s,0}(\eta;\bm{\lambda})
  &=A\times\Xi_{d_1\ldots d_s}(\eta;\bm{\lambda}+\bm{\delta}),
  \label{plusdelta}\\
  A&=\left\{
  \begin{array}{ll}
  (-1)^{s_{\I}}\prod\limits_{j=1}^{s_{\II}}
  (g-d_j-\tfrac12)&:\text{L}\\
  2^{-s_{\I}}\prod\limits_{j=1}^{s_{\I}}(h-d_j-\tfrac12)\cdot
  (-2)^{-s_{\II}}\prod\limits_{j=1}^{s_{\II}}(g-d_j-\tfrac12)&:\text{J}
  \end{array}\right.,\nonumber
\end{align}
where $j$ runs for type $\I$ $d_j$ (or type $\II$ $d_j$) in the products
$\prod\limits_{j=1}^{s_{\I}}$ (or $\prod\limits_{j=1}^{s_{\II}}$).
Therefore the ground state has the form,
\begin{equation}
  \phi_{d_1\ldots d_s\,0}(x;\bm{\lambda})\propto
  \phi_{0}(x;\bm{\lambda}^{[s_{\I},s_{\II}]})
  \frac{\Xi_{d_1\ldots d_s}\bigl(\eta(x);\bm{\lambda}+\bm{\delta}\bigr)}
  {\Xi_{d_1\ldots d_s}\bigl(\eta(x);\bm{\lambda}\bigr)}.
  \label{phis0}
\end{equation}
Then eqs.\eqref{Fsdef}--\eqref{Bsdef} become
\begin{align}
  \mathcal{F}_{d_1\ldots d_s}(\bm{\lambda})
  &=\cF\frac{\Xi_{d_1\ldots d_s}(\eta;\bm{\lambda}+\bm{\delta})}
  {\Xi_{d_1\ldots d_s}(\eta;\bm{\lambda})}\Bigl(\frac{d}{d\eta}
  -\frac{\partial_{\eta}\Xi_{d_1\ldots d_s}(\eta;\bm{\lambda}+\bm{\delta})}
  {\Xi_{d_1\ldots d_s}(\eta;\bm{\lambda}+\bm{\delta})}\Bigr),
  \label{Fs}\\
  \mathcal{B}_{d_1\ldots d_s}(\bm{\lambda})
  &=-4\cF^{-1}c_2(\eta)\frac{\Xi_{d_1\ldots d_s}(\eta;\bm{\lambda})}
  {\Xi_{d_1\ldots d_s}(\eta;\bm{\lambda}+\bm{\delta})}
  \Bigl(\frac{d}{d\eta}
  +\frac{c_1(\eta,\bm{\lambda}^{[s_{\I},s_{\II}]})}{c_2(\eta)}
  -\frac{\partial_{\eta}\Xi_{d_1\ldots d_s}(\eta;\bm{\lambda})}
  {\Xi_{d_1\ldots d_s}(\eta;\bm{\lambda})}\Bigr),
  \label{Bs}
\end{align}
where the functions $c_1(\eta;\bm{\lambda})$ and $c_2(\eta)$ are those
appearing in the (confluent) hypergeometric equations for the Laguerre and
Jacobi polynomials
\begin{equation}
  c_1(\eta,\bm{\lambda})\eqdef\left\{
  \begin{array}{ll}
  g+\tfrac12-\eta&:\text{L}\\
  h-g-(g+h+1)\eta&:\text{J}
  \end{array}\right.,\quad
  c_2(\eta)\eqdef\left\{
  \begin{array}{ll}
  \eta&:\text{L}\\
  1-\eta^2&:\text{J}
  \end{array}\right..
  \label{c1c2}
\end{equation}
The second order differential operator
$\widetilde{\mathcal{H}}_{d_1\ldots d_s}(\bm{\lambda})$ governing the
multi-indexed polynomials is:
\begin{align}
  &\widetilde{\mathcal{H}}_{d_1\ldots d_s}(\bm{\lambda})
  \eqdef\Psi_{d_1\ldots d_s}^{\,-1}(x;\bm{\lambda})\circ
  \mathcal{H}_{d_1\ldots d_s}(\bm{\lambda})\circ
  \Psi_{d_1\ldots d_s}(x;\bm{\lambda})
  =\mathcal{B}_{d_1\ldots d_s}(\bm{\lambda})
  \mathcal{F}_{d_1\ldots d_s}(\bm{\lambda})\n
  &\phantom{\widetilde{\mathcal{H}}_{d_1\ldots d_s}(\bm{\lambda})}
  =-4\biggl(c_2(\eta)\frac{d^2}{d\eta^2}
  +\Bigl(c_1(\eta,\bm{\lambda}^{[s_{\I},s_{\II}]})-2c_2(\eta)
  \frac{\partial_{\eta}\Xi_{d_1\ldots d_s}(\eta;\bm{\lambda})}
  {\Xi_{d_1\ldots d_s}(\eta;\bm{\lambda})}\Bigr)\frac{d}{d\eta}\n
  &\phantom{\widetilde{\mathcal{H}}_{d_1\ldots d_s}(\bm{\lambda})}
  \qquad\quad
  +c_2(\eta)
  \frac{\partial^2_{\eta}\Xi_{d_1\ldots d_s}(\eta;\bm{\lambda})}
  {\Xi_{d_1\ldots d_s}(\eta;\bm{\lambda})}
  -c_1(\eta,\bm{\lambda}^{[s_{\I},s_{\II}]}-\bm{\delta})
  \frac{\partial_{\eta}\Xi_{d_1\ldots d_s}(\eta;\bm{\lambda})}
  {\Xi_{d_1\ldots d_s}(\eta;\bm{\lambda})}
  \biggr),
  \label{Hts}\\
  &\widetilde{\mathcal{H}}_{d_1\ldots d_s}(\bm{\lambda})
  P_{d_1\ldots d_s,n}(\eta;\bm{\lambda})=\mathcal{E}_n(\bm{\lambda})
  P_{d_1\ldots d_s,n}(\eta;\bm{\lambda}).
  \label{HtsPsn}
\end{align}

For appropriate parameter range (see \cite{os25}), the operators
$\mathcal{H}_{d_1\ldots d_s}$, $\hat{\mathcal{A}}_{d_1\ldots d_s}$, 
$\mathcal{A}_{d_1\ldots d_s}$, etc.\! are non-singular, and we have the
norm formula,
$(\phi_{d_1\ldots d_s\,n},\phi_{d_1\ldots d_s\,m})
=\prod\limits_{j=1}^s(\mathcal{E}_n-\tilde{\mathcal{E}}_{d_j})
\cdot(\phi_n,\phi_m)$ with $(f,g)\eqdef\int_{x_1}^{x_2}dxf(x)g(x)$.
For equivalence among the multi-indexed polynomials,
see \cite{equiv_miop,t13}.


\end{document}